\documentclass[12pt]{article}
\usepackage{codams,float,amssymb,amsmath}
\usepackage{graphicx}
\usepackage[bookmarks,bookmarksopen,
  pdftitle={"Nodally 3-connected planar graphs and convex combination mappings"},
  pdfauthor={Colm \'O D\'unlaing}]{hyperref}

\newtheorem{definition}[theorem]{Definition}
\newtheorem{lemma}[theorem]{Lemma}

\newtheorem{proposition}[theorem]{Proposition}
\newtheorem{corollary}[theorem]{Corollary}

\numberwithin{equation}{section}

\renewcommand{\int}{\text{\rm interior}}
\newcommand{\ext}{\text{\rm exterior}}
\newcommand{\hull}{\text{\rm hull}}
\newcommand{\be}{\begin{equation*}}
\newcommand{\ee}{\end{equation*}}
\DeclareGraphicsExtensions{.eps}
\title{Nodally 3-connected planar graphs and convex combination mappings}
\author{Colm \'O D\'unlaing\\
{\texttt{ odunlain@maths.tcd.ie}}\\
{\em Mathematics, Trinity College, Dublin 2, Ireland}\thanks{
These results were presented at the fourth
Irish MFCSIT conference, Cork, Ireland, August 2006.}}
\begin{document}
\maketitle

\begin{abstract}
A barycentric mapping of a planar graph is a plane
embedding in which every internal vertex is the average
of its neighbours.  A celebrated result of Tutte's \cite{tutte}
is that if a planar graph is nodally 3-connected then
such a mapping is an embedding.  Floater generalised
this result to convex combination mappings in which every
internal vertex is a proper weighted average of its neighbours.
He also generalised the result to all triangulated
planar graphs.

This has applications in numerical analysis (grid generation),
and in computer graphics (image morphing, surface triangulations,
texture mapping): see \cite{floater02,white}.

White \cite{white} showed that every chord-free triangulated planar
graph is nodally 3-connected.

We show that (i) a nontrivial plane embedded graph is nodally 3-connected if
and only if every face boundary is a simple cycle and
the intersection of every two faces is connected;
(ii) every convex combination mapping of a plane embedded graph
$G$
is an embedding if and only if (a) every face boundary
is a simple cycle, (b) the intersection of every two {\em bounded}
faces is connected, and (c) there are no so-called inverted
subgraphs; (iii) this is equivalent to $G$ admitting a convex
embedding (see \cite{stein}); and (iv) any two such
embeddings (with the same orientation) are isotopic.
\end{abstract}

\section{Planar graphs and nodal 3-connectivity}
\label{criterion} 
We follow the usual definitions of graphs, including paths, simple paths,
cycles, simple cycles, and connectivity:
\cite{kant} is a useful source on the subject.
The accepted definition of graph does not allow self-loops nor multiple edges
nor infinite sets of vertices,
so it is a finite simple graph in Tutte's language \cite{tutte},
and a graph $G$ can be specified as a pair $(V,E)$ giving
its vertices and edges. $E$ is a set of
unordered pairs of distinct vertices in $V$.  Two vertices
$u,v$ are {\em adjacent} or {\em neighbours} if
$\{u,v\} \in E$.

Given $G = (V,E)$,
when $u$ is considered to be a  vertex, $u\in G$ means $u\in V$,
and when $e$ is considered to be an edge, $e\in G$ means $e\in E$.

\numpara
\label{subgraphs etcetera} 
{\bf Subgraphs, etcetera.}
Given $G=(V,E)$ and $G' = (V',E')$, $G'$ is a {\em subgraph}
of $G$ if $V'\subseteq V$ and $E'\subseteq E.$

Given $G$ and given $S \subseteq V$, the {\em subgraph of $G$
spanned by $S$} is the graph $(S,E')$ where
$$ E' = \{ \{u,v\}\in E:~~ u,v \in S\}.$$

The {\em degree} (in $G$) $\deg(v)$ of a vertex $v$ is the number of edges
incident to it, or the number of neighbours it has.
The word `node' is reserved in \cite{tutte} to denote
vertices whose degree $\not=$ $2$.

A {\em path} in $G$ is a sequence $u_0, \ldots, u_k$ of vertices
where $k\geq 0$ and for $0 \leq j \leq k-1$, $\{u_j, u_{j+1} \} \in E$.
It is {\em simple} if all the vertices $u_j$ are distinct.
The {\em inner vertices} in a simple path are
$\{u_1,\ldots,u_{k-1}\}$.

A {\em cycle} is a path $u_0,\ldots, u_k,u_0$ (that is, its
first and last vertices are the same).
It is a {\em simple cycle} if $k=0$ or the path $u_0,\ldots,u_k$
is a simple path.

If we write, say, $v_1,\ldots, v_n$ for a cycle, it is implied
that $v_n$ is the second-last vertex rather than a recurrence
of the first, so properly the cycle is $v_1,\ldots, v_n,v_1$.

If $G_i = (V_i,E_i)$ are
two graphs then we define 
$$G_1 \cap G_2 = (V_1\cap V_2, E_1\cap E_2)\quad\text{and}\quad
G_1 \cup G_2 = (V_1\cup V_2, E_1\cup E_2).$$%
If $G = (V,E)$ and $S \subseteq V$ then
$ G\backslash S = (V',E')$ where
$$ V' = V \backslash S\quad\text{and}\quad
E' = \{ \{u,v\} \in E:~ u\notin S ~\text{and}~ v\notin S\}.$$ We extend
this notation loosely but with little risk
of confusion: if $x$ is a vertex then $G\backslash x = 
G\backslash \{x\}$, and if $H$ is a subgraph, or a path, or a cycle,
then $G\backslash H$ is the same as $G\backslash S$ where $S$ is the set
of vertices in $H$.

$G$ is {\em connected} if every two vertices are connected
by a path in $G$.
$G$ is {\em biconnected} if it is
connected and for every $u\in G$, $G\backslash u$ is connected.
$G$ is {\em triconnected} if it is biconnected and
for any $u,v\in G$, $G\backslash \{u,v\}$ is connected.
(Here $\{u,v\}$ is a pair of vertices, not necessarily
an edge.)

A {\em path (graph)} is either a trivial graph or
a connected graph in which two vertices have degree $1$ and all
others have degree $2$.
A {\em simple cycle (graph)} is a connected nonempty graph all of
whose vertices have degree $2$.

This paper is concerned with {\em nodal 3-connectivity}
(defined in \ref{nodal 3-connectivity}),
which requires biconnectivity but is weaker
than triconnectivity.

\begin{definition}
\label{planar graph defined} 
Let $G = (V,E)$ be a graph.
\begin{itemize}
\item
The unit interval $\{t \in \IR:~ 0 \leq t \leq 1\}$
is denoted $[0,1]$.
Given distinct points $x$ and $y$ in $\IR^2$,
a {\em simple curve-segment} joining $x$ to $y$ is
continuous, injective map $\pi: [0,1] \to \IR^2$
such that $\pi(0) = x$ and $\pi(1) = y.$
\item
Let $f$ be a map taking
each vertex $u$ to a point $f(u)$ in the plane $\IR^2$,
and each edge $e=\{u,v\}$ to a simple curve-segment
$f(e)$ joining $f(u)$ to $f(v)$.

The {\em relative interior} of $e$, which depends on $f$,
is the open curve-segment
$$ \int(e) = f(e) \backslash \{f(u)\} \backslash \{f(v)\}.$$

\item
The map $f$ is a {\em plane embedding} of $G$ if the points
$f(u)$ are distinct and the relative interiors of any two
edges are disjoint.

\item
A plane embedding $f$ is {\em straight-edge} if $f(e)$ is
a line-segment for every edge $e$.
\item
$G$ is {\em planar} if a plane embedding exists.
\end{itemize}
\end{definition}

\begin{figure}
\centerline{
\includegraphics[height=1in]{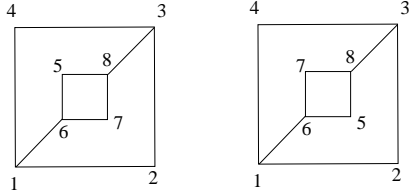}}
\caption{a graph with different plane embeddings. Also, the
barycentric map is not an embedding.}
\label{nontutte.fig}
\end{figure}

One often speaks of a planar graph $G$ with a specific
plane embedding of $G$ in mind, so it really means a
plane embedded graph.  A very significant difference is
that a plane embedded graph has a definite external face
(Definition \ref{faces of graph}),
whereas there is no notion of external face, nor
perhaps even of face, in a planar graph without a prescribed
embedding.
Figure \ref{nontutte.fig} shows a planar graph with two
quite different embeddings.

Plane embeddings could somehow be pathological
and they should be discussed in terms of
the Jordan Curve Theorem mentioned below.
However, the following proposition could be used
to simplify the arguments.

\begin{proposition}
\label{straight-edge embeddings} 
Every planar graph
admits a straight-edge
embedding {\rm \cite{dpp,od94,read}}.\qed
\end{proposition}

\numpara
\label{topology in two dimensions} 
{\bf Topology in two dimensions.} See \cite{moise,stillwell}.
We assume the basic notions
of open and closed sets, connectedness, and path-connectedness.
If $x\in \IR^2$ and $\varepsilon > 0$ then the
$\varepsilon$-neighbourhood of $x$ is
$$ B(x,\varepsilon) = \{ y \in \IR^2:~~ |y-x| < \varepsilon\}.$$
If $S$ is any subset of $\IR^2$ then
its {\em closure}, written $\overline{S}$, is
$$ \overline{S} =
\{ x\in\IR^2:~~ (\forall \varepsilon>0)
B(x,\varepsilon)\cap S \not=\emptyset\},$$ and
its {\em boundary} $\partial S$ is
$$\partial S = \overline{S}\cap\overline{\IR^2\backslash S}.$$ If $S$
is open then $S\cap \partial S = \emptyset$. We
are not concerned with connectedness, but with the
rather stronger notion of path-connectedness:
a set $S$
is {\em path-connected} if for any $x,y\in S$
there exists a path from $x$ to $y$, a continuous map $\pi:[0,1]\to S$
such that $\pi(0) = x$ and $\pi(1) = y$.

\numpara
\label{Jordan curves} 
{\bf Jordan curves.} A {\em Jordan curve} is a subset of
$\IR^2$ homeomorphic to the unit circle $S^1$.
That is, $J$ is a Jordan
curve iff there exists a continuous injective map $h: S^1 \to \IR^2$
whose range is the set $J$.

\begin{proposition}
\label{path components in graph and plane} 
Let $x$ and $y$ be two vertices in a plane embedding $f$ of
a graph $G$.  Then they are in the same component
of $G$ as a graph if and only if they are in
the same path-component of $G$ as a topological
subspace of $\IR^2$. Also if $C$ is a simple cycle then
its image under
$f$ is a Jordan curve.  (Proof easy.)\qed
\end{proposition}

Part (i) of Proposition \ref{jordan curve theorem} below
states the Jordan Curve Theorem, which is a difficult result.
Proofs usually involve
algebraic topology \cite{greenberg}, but less advanced
methods can be used
\cite{moise,stillwell}.
Actually for our purposes we need only consider polygonal
Jordan curves, which makes the proofs much easier.
Part (ii) is elementary.

\begin{proposition}
\label{jordan curve theorem} 
{\rm (i)} (Jordan Curve Theorem {\rm \cite{greenberg,moise,stillwell}}).
If $J$ is a Jordan curve
then $\IR^2 \backslash J$ is the union of two
open, path-connected components, 
$\int(J)$ and $\ext(J)$,
$\int(J)$, the {\em inside}, is bounded, and $\ext(J)$, the {\em outside}
or {\em exterior},
is unbounded, and $\partial (\int(J)) = \partial (\ext(J)) = J.$

{\rm (ii)} If $S$ is any path-connected open set such that
$\partial S = J$, then $S=\int(J)$ or $S=\ext(J)$.\qed
\end{proposition}

\numpara
\label{edges inside and outside Jordan curves} 
{\bf Edges inside and outside Jordan curves.}
If $J$ is a Jordan curve and $e=\{u,v\}$ an edge of a
graph, and $f$ an embedding such that $f(e)$ doesn't meet $J$
except perhaps at $f(u)$ or $f(v)$, then
the relative interior of $e$ (Definition
\ref{planar graph defined}) satisfies
$$ \int(e) \subseteq \int(C)
\quad\text{or}\quad
\int(e) \subseteq \ext(C).$$

In this case we say $e$ is inside or outside $J$ as appropriate.
In Section \ref{ambient isotopy section}
we shall need a certain refinement of the Jordan curve theorem:

\begin{proposition}
\label{schoenflies theorem} 
{\bf (Jordan-Sch\"onflies Theorem).}
Let $D^1$ be the unit disc in $\IR^2$ and $S^1 = \partial D^1,$
the unit circle.  Then if $J$ is a Jordan curve (a homeomorphic
image of $\partial D^1$), the homeomorphism of $\partial D^1$
extends to a homeomorphism between $D^1$ and $\overline{\int(J)}.$

More generally, if $J$ and $J'$ are two Jordan curves then
the homeomorphism between $J$ and $J'$ extends to a homeomorphism
between $\IR^2$ and itself taking $\int(J)$ to $\int(J')$
and $\ext(J)$ to $\ext(J').$
(See {\rm \cite{moise}}.)\qed
\end{proposition}

\begin{definition}
\label{faces of graph} 
Given a plane embedding $f$ of a graph $G$, by abuse of notation
let $G$ also denote the union of points and curve-segments
constituting its image in the plane.  This is a closed and
bounded set of points in the plane.

A {\em face} of $G$ is a path-connected component
of $\IR^2\backslash G$.

All faces except one are bounded.  The unbounded face
is called the {\em external face} or {\em outer face}.  Vertices on the
external face  are called {\em external}; the others are
{\em internal}.

The plane embedding is {\em triangulated} if every bounded
face is incident to exactly three edges, and {\em fully triangulated}
if every face, bounded and unbounded, is incident to three edges.
\end{definition}

Faces are open sets in $\IR^2$.

\begin{definition}
\label{triangulation} 
Let $f$ be a plane embedding of a graph
$G = (V,E)$. A {\em triangulation} of the
graph is a triangulated plane embedding $f'$ of
a graph $G' = (V',E')$ where $V' = V$ and
$E' \supseteq E$, where $f'(u) = f(u)$ for all
$u\in V$ and $f'(e) = f(e)$ for all $e \in E$.
\end{definition}

\begin{proposition}
\label{all embeddings triangulable} 
Every plane embedded graph can be triangulated
{\rm \cite{kant}}.\qed
\end{proposition}

\begin{proposition}
\label{partial F is a subgraph} 
{\rm (i)} If $F$ is a face of a plane embedded graph $G$,
then $\partial F$ is a subgraph of $G$, and
{\rm (ii)} $G=\bigcup_F \partial F$. (Proof omitted.)\qed
\end{proposition}

\numpara
\label{convexity} 
{\bf Convex sets in the plane.}
We note the basic definitions and results (see \cite{brondstred}).
A set $A$ is {\em convex} if for any two points $a,b \in A$,
the line-segment $ab$ is entirely contained in $A$.  Suppose $S$ is
a finite set of points in the plane.
The {\em convex hull} $\hull(S)$
is the smallest convex set containing $S$,
that is, the intersection of all convex sets containing $S$.
It is also the intersection of all closed half-planes containing
$S$.  Either $\hull(S)$ is empty, or a point, or a
line-segment, or it is bounded by a convex polygon
whose corners are in $S$.  In the latter case $\hull(S)$ is
the intersection of those closed half-planes containing
$S$ whose boundaries contain sides of $S$.

\begin{proposition}
\label{closure convex} 
If $A$ is convex then its closure
$\overline{A}$ is convex. (Proof easy.)\qed
\end{proposition}

\begin{definition}
\label{convex combination map} 
{\bf (convex combination maps)} {\rm \cite{floater03}}.
A {\em convex embedding} of a planar graph $G$ is a straight-edge
embedding
in which all bounded faces are convex, and the outer boundary
is a simple polygon.

\begin{figure}
\centerline{
\includegraphics[width=2in]{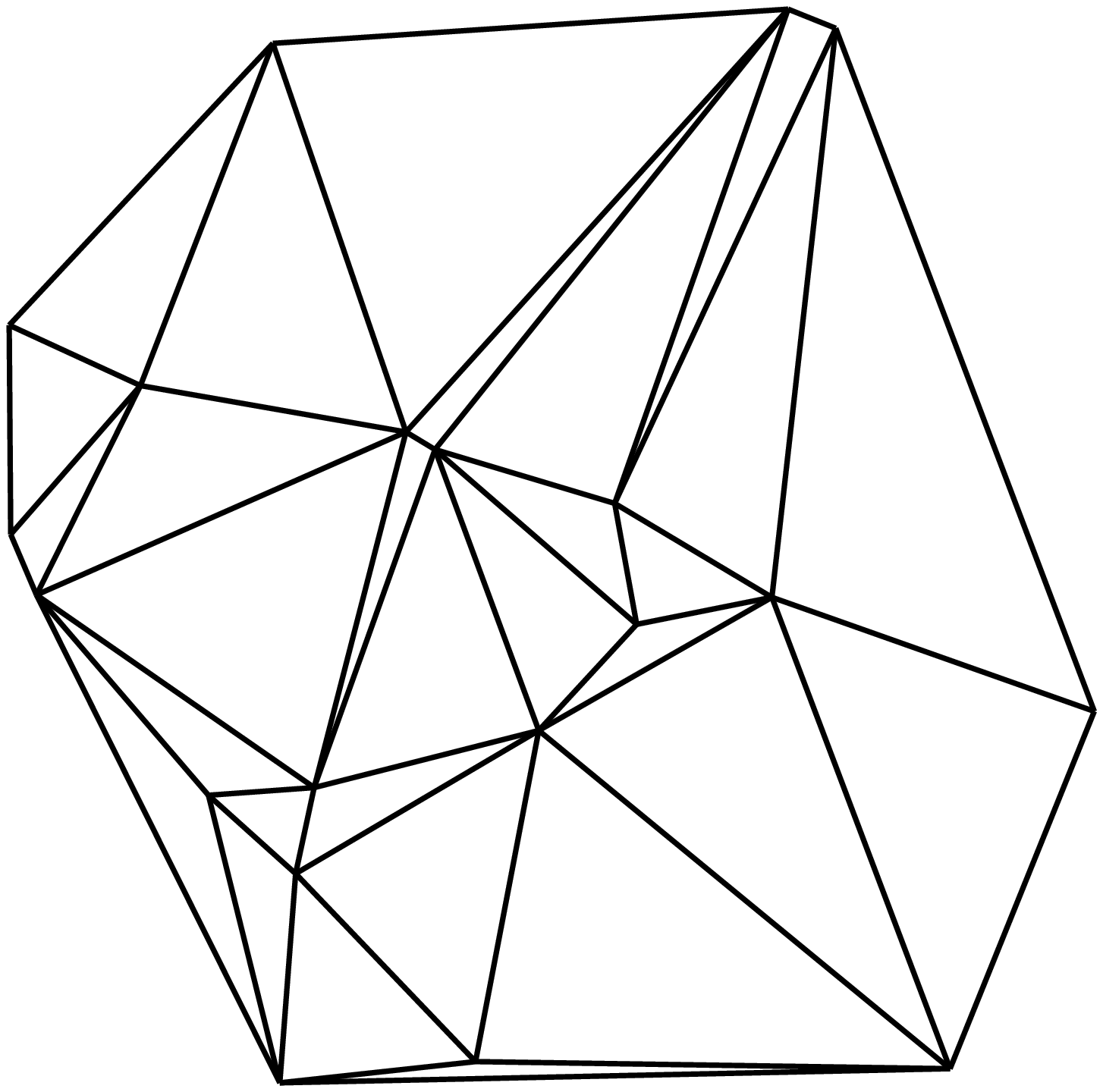}\hspace*{.5in}
\includegraphics[width=2in]{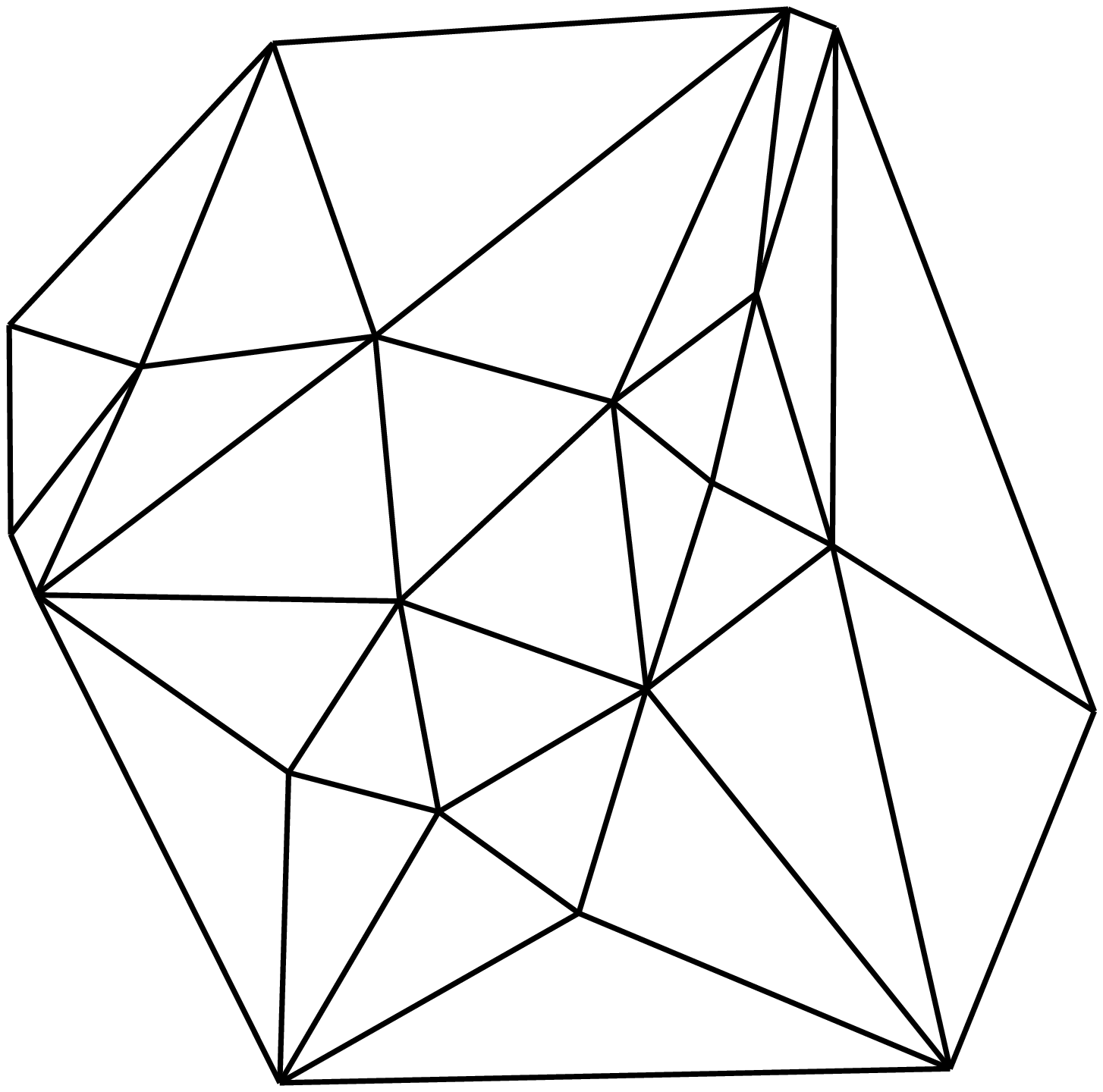}\hspace*{.5in}\ \ 
}
\caption{Delaunay triangulation of 20 points and barycentric
embedding of the same graph with the same bounding
polygon.}
\label{delaun20.fig}
\end{figure}

Let $G$ be a plane embedded graph whose external boundary is
a simple cycle $C$.
Another map $f$ from its vertices to
points in the plane is {\em a convex combination map} if
{\rm (a)} there exist coefficients $\lambda_{uv}$ ($u,v$ vertices) such that
\begin{itemize}
\item
$\lambda_{uv} \geq 0$,
and
$\sum_v \lambda_{uv} = 1$.
\item
If $v$ is an external vertex then $\lambda_{vv} = 1.$
\item
If $u$ and $v$ are adjacent and $u$ is internal
then $\lambda_{uv} > 0$.
\item
Otherwise $\lambda_{uv} = 0$.
\end{itemize}

{\rm (b)}
the external
vertices are mapped (in cyclic order) to the corners
of a convex polygon, and {\rm (c)} for
every internal vertex $u$, that is, for every vertex $u\notin C$,
\begin{equation}
\label{weighted average of neighbours} 
 f(u) =  \sum_v \lambda_{uv} f(v)
\end{equation}

The map is a {\em barycentric map} if
for each internal vertex $u$ and neighbour $v$ of
$u$, $\lambda_{uv} = 1/\deg(u)$.
If a barycentric map determines a straight-edge embedding
of $G$ then it is called a {\em barycentric embedding}.
\end{definition}

For example, Figure \ref{delaun20.fig} shows a Delaunay triangulation
with 20 vertices, and a barycentric embedding of the same graph.

The definition of convex embedding does not exclude the
possibility that several edges on a face boundary be collinear.
Tutte's definition of convex embedding
\cite{tutte60} requires that the external boundary
be a convex polygon, which would rule out most triangulated
graphs.  Hence we require that it be a simple
polygon, though not necessarily convex.

In a barycentric map, every internal vertex is the average, centroid,
or barycentre, of its neighbours.  In a convex combination map
every internal vertex is a proper weighted average of its
neighbours.

The following simple lemma is very useful.

\begin{lemma}
\label{lem: one-sided}
Let $f$ be a convex combination map, $H$ a closed convex
set, and $v$ an internal vertex such that
for all neighbours $u$ of $v$, $f(u)\in H$.
If, for some neighbour $u$ of $v$, $f(u)\in H^o$ (the topological interior
of $H$), then $v \in H^o$.
\end{lemma}

{\bf Proof.} Fix a neighbour $u$ such that $f(u)\in H^o$,
and fix $\varepsilon > 0$ so for all points $x$ in the
plane, if $|x|<\varepsilon$, then $x+f(u)\in H$.

Since $v$ is internal,
$$ f(v) = \sum_w \lambda_{vw} f(w),$$  and $f(v)\in H$.
The sum can be written
as $\lambda_{vu} f(u) + (1-\lambda_{vu})y$  where $y$ is
a proper weighted average of the other neighbours of $v$ --- or
$O$ if $\lambda_{vu}=1$.

Since $H$ is convex,
$$ \{ \lambda_{vu}(x+f(u)) + (1-\lambda_{vu})y:~ |x| < \varepsilon \}
\subseteq H.$$ This is the open disc around $f(v)$ of
radius $\lambda_{vu}\varepsilon$, so $f(v)\in H^o$. {\bf Q.E.D.}\medskip

\begin{lemma}
\label{convex combination map inside P} 
If $f$ is a convex combination map taking the external boundary
of a connected plane embedded graph $G$ to a convex polygon $P$,
then all vertices and edges are mapped by $f$ into $\hull(P)$.
\end{lemma}

{\bf Proof.}
Let $D=\hull(P)$.
Since $D$ is convex, it is enough to show that for every
vertex $u$, $f(u)\in D.$  External vertices are mapped
to corners of $P,$ hence into $D.$

Suppose there is an internal vertex $w$ such that
$f(w)\notin D.$  $D$ is the intersection of finitely many
closed half-planes, and one of them
does not contain $f(w).$  By changing coordinates if
necessary, it can be arranged that $D$ is bounded above
by the $x$-axis and there exist vertices $u$
such that $f(u)$ is above the $x$-axis.
Choose $u$ so $f(u)$ has maximal $y$-coordinate, $h$, say, and
let $H$ be the close half-plane $y \leq h$.

Since $G$ is connected, there is a path
$$ u_0, \ldots , u_k = u $$ where $u_0$ is an external
vertex.  Since $f(u_0) \in D$, $f(u_0)$ is in the interior $H^o$
of $H$,
so without loss of generality, $f(u_{k-1})\in H^o$ and
by Lemma \ref{lem: one-sided}, $f(u)\in H^o$, a contradiction.
{\bf Q.E.D.}\medskip

\begin{lemma}
\label{lem: conv comb emb conv faces}
If a convex combination map is an embedding, then its
embedded faces are convex.
\end{lemma}

{\bf Proof.} Let $F$ be a bounded face.
Since $f$ is a straight-edge embedding, $f(\partial F)$
is a simple polygon, and we need only show it has
no concave corners.  However, if $f(v)$ is a concave
corner then $v$ is an inner vertex and
there is a convex wedge $V$ such that $f(u)\in V$
for all neighbours $u$ of $v$. Let $H$ be a closed
half-plane such that $V\subseteq H$ and $V\backslash H^o
=\{f(v)\}$. By Lemma \ref{lem: one-sided}, $f(v)\in H^o$, a contradiction.
{\bf Q.E.D.}\medskip

\numpara
\label{matrix defining a convex combination map} 
{\bf Matrix defining a convex combination map.} Given
a plane embedded graph $G$ whose external boundary
is a simple cycle $C$, convex combination maps are easily specified
using a matrix $A$.  Suppose that $G$ has
$m$ vertices $v_1,\ldots,v_m$, the first $n$ of them
belonging to $C$, the last $m-n$ being
internal vertices, and the coordinates of their
images are $x_i, y_i, 1 \leq i \leq m$.  Any map from
vertices to points, including any straight-edge
embedding, is equivalent to a column vector of height $2m$.

Let $A$ be the $m\times m$ matrix whose first
$n$ rows are identical with those of the identity
matrix, and whose
last $m-n$ rows express the barycentric mapping
equations (\ref{weighted average of neighbours}). Equivalently,
for $1 \leq i,j \leq m$, let
$$ a_{ij} = \begin{cases}
1\quad\text{if}~i=j,\\
0\quad\text{if}~ i \not= j ~\text{and}~ j \leq n,~\text{and}\\
-\lambda_{v_iv_j} \text{if} ~ i \not= j.
\end{cases}
$$ Equation
\ref{weighted average of neighbours} can be written in
the form
$$ \sum a_{ij} x_j = 0\quad\text{and}\quad \sum a_{ij}y_j = 0,
\quad ( n < i \leq m).$$ For
any convex combination map $f$ (with
$\lambda_{uv}$ given), let $B_x$ be the column vector of
height $m$
whose first $n$ entries give the $x$-coordinates
of the corners of $P$
and whose other entries are zero; similarly
let $B_y$ specify the $y$-coordinates. Then $f$ is
equivalent to column vectors $X$ and $Y$ satisfying
$$ AX = B_x;\quad AY = B_y.$$

\begin{lemma}
\label{unique convex combination map} 
{\rm (i)} If $G$ is connected then the above matrix $A$ is
invertible.

{\rm (ii)} If $G$ is a connected plane embedded graph whose external
boundary is a simple cycle, and whose external
vertices are mapped in cyclic order to the corners of
a convex polygon, and weights $\lambda_{uv}$ are given,
then this map extends to a unique
convex combination map of $G$.
\end{lemma}

{\bf Sketch of proof.}
(See \cite{tutte,bsst,floater97,white}.)
Tutte's proof of (i) \cite{tutte,bsst} says that
the determinant of $A$ (scaled up) is the number of spanning trees of
a certain connected graph related to $G$.
There is a much more transparent proof given
in \cite{floater97} and also in \cite{white} saying
that if $A$ has nonzero kernel then one can follow
a path from an external vertex to an internal vertex
where the internal vertex cannot satisfy Equation
\ref{weighted average of neighbours}.
Part (ii) follows trivially.\qed

\begin{definition}
\label{nodal 3-connectivity} 
A graph $G$ is {\em nodally 3-connected} if it
is biconnected and for every two subgraphs
$H$ and $K$ of $G$, if $G = H\cup K$ and
$H\cap K$ consists of just two vertices (and
no edges), then $H$ or $K$ is a simple path.
\end{definition}

\begin{proposition}
\label{nodally 3-connected and no deg 2} 
Every triconnected graph is nodally 3-connected,
and every nodally 3-connected graph with no vertices
of degree 2 is triconnected. (Proof omitted.)\qed
\end{proposition}

\begin{definition}
\label{peripheral polygon} 
A {\em peripheral polygon} in a connected graph $G$
is a simple cycle $C$ such that $G\backslash C$ is
connected.
\end{definition}

The following result of Tutte's is fundamental.

\begin{proposition}
\label{tutte's theorem} 
{\em (Tutte \cite{tutte})}. If $G$ is a nodally 3-connected
planar graph\footnote{with a few
exceptions: see Figure \ref{nontutte2.fig}. The result is phrased
differently in \cite{tutte}.}  and $C$ is a peripheral polygon, and
the vertices of $C$ are mapped (in cyclic order) onto the corners of a convex
polygon $P$, then that map extends to a unique barycentric map
which is a convex, straight-edge embedding of $G$.\qed
\end{proposition}

It is easy to give a counterexample when $G$ is not
nodally 3-connected.  For example, in Figure \ref{nontutte.fig},
any barycentric map must map
the inner square face to a line-segment.  The figure
illustrates different plane embeddings of the same
graph, which is not nodally 3-connected.

We shall rely more heavily on the following

\begin{proposition}
\label{floaters theorem} 
{\em (Floater \cite{floater03}).}
If $G$ is a triangulated (plane embedded) graph, then every convex
combination map of $G$ is an embedding.\qed
\end{proposition}

Theorem \ref{planar nodally 3-connected} below shows that,
except regarding the external face, a {\em planar} graph
is nodally 3-connected if and only if barycentric maps
are plane embeddings.

%
%

Lemmas
\ref{plane disconnected face disconnected} and
\ref{planar biconnected} below are fairly obvious
and well-known, but still worth mentioning.

\begin{lemma}
\label{plane disconnected face disconnected} 
A plane embedded graph $G$ is
connected if and only if for every face $F$,
the boundary $\partial F$ is (path-)connected.\qed
\end{lemma}

%

\begin{proposition}
\label{eulers formula} 
{\bf (Euler's Formula.)}
If $G$ is a plane (straight-edge) embedded graph
then 
$$v-e+f = c+1,$$ where $v,e,f$, and $c$ are the
numbers of vertices, edges, faces, and components
of $G$. (Proof omitted.)\qed
\end{proposition}

\begin{figure}
\centerline{\includegraphics[height=1.5in]{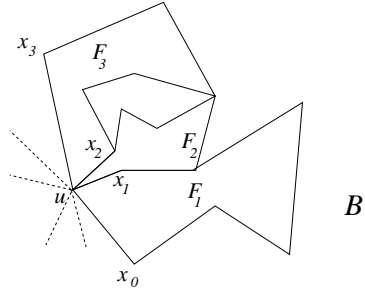}}
\caption{neighbours of $u$ connected by paths avoiding $u$.}
\label{B.fig}
\end{figure}

\begin{lemma}
\label{link of u} 
Let $G$ be a straight-edge embedded plane graph in which all face boundaries
are simple cycles, and let $u$ be any vertex of $G$.

Let $x_0, \ldots , x_k$ be a list of neighbours of
$u$ consecutive in anticlockwise order; possibly $x_0 = x_k$
but otherwise they are distinct.  For $1 \leq j \leq k$
let $F_j$ be the face occurring between the edges (line-segments)
$u x_{j-1}$ and $u x_j$ in the anticlockwise sense.  (The
faces $F_j$ are not necessarily distinct.)

Let $B$ be the subgraph formed by the edges and vertices
in $\bigcup_j \partial F_j$.

Then any two vertices in the list $x_j$ are joined by a path in
$B \backslash u$. See Figure \ref{B.fig}.
\end{lemma}

{\bf Proof.}
$B \backslash u$ is also the subgraph consisting
of all vertices and edges in $\bigcup_j (\partial F_j \backslash u)$.
Since each face is a simple cycle, $\partial F_j \backslash u$
is a path joining $x_{j-1}$ to $x_j$.  Thus
$B \backslash u$ contains paths joining all these vertices
$x_j$. {\bf Q.E.D.}\medskip

\begin{lemma}
\label{planar biconnected} 
A plane straight-edge embedded graph $G$ is biconnected if and only if
the graph consists of a single vertex or a single edge, or
the boundary of every face is a simple cycle.
\end{lemma}

{\bf Sketch proof.}
(i): If. A single vertex or edge is biconnected, so we
assume that the boundary of every face is a simple cycle.
$G$ is connected (Lemma
\ref{plane disconnected face disconnected}).

For any vertex $x$ and all neighbours $x_j$ of $x$ there
exist paths connecting these neighbours which avoid $x$
(Lemma \ref{link of u}).  Therefore all these neighbours are
in the same component of $G\backslash x$, and it follows that
$G\backslash x$ is connected.  Hence $G$ is biconnected.

(ii): Only if. Suppose that $G$ is connected, not a single vertex
or edge, and there exists a face $F$ whose
boundary is not a simple cycle (graph):
$\partial F$ is connected but
contains a node $x$ whose degree (in $\partial F$, not in $G$)
differs from $2$.  If $\partial F$ contained a vertex  of
degree $0$ then (since $G$ is nontrivial) $G$ would be disconnected.
If it contained a vertex of degree $1$, then $G$ would
be disconnected or not biconnected.  Hence we can assume
that all vertices on $\partial F$ have degree $\geq 2$ in $\partial F$.

Let $u\in \partial F$ be a vertex of degree $\geq 3$ in $\partial F$.  Let
$x_1,\ldots , x_k$ be the vertices adjacent to $u$ in anticlockwise
order.  For $1 \leq j\leq k$, $x_j u x_{j+1}$ ($x_{k+1}=x_1)$ forms a clockwise
part of the boundary of a face incident to $u$.  Since $u$ has
degree $\geq 3$ in $\partial F$, at least two of these
paths are incident to $F$ and there are fewer than $k$
distinct faces incident to $u.$

Let $G' = G\backslash \{u\}$. All faces incident to $u$ in $G$
merge into a single face of $G',$ and the other faces of $G$
are preserved.
The Euler formula gives
$$ v - e + f = 2$$ for $G$, since $G$ is connected.  Correspondingly
for $G'$,
$$ v' - e' + f' = 1 + c'.$$ Now $v' = v-1,$ and $e' = e - k$.
Since in $G'$ fewer than $k$ faces are merged into a single face,
$f' > f+1-k$. Therefore
$$ v' - e' + f' > v-1 - e + k + f + 1 - k = 2,$$ so $c' > 1$,
$G'$ is disconnected, and $G$ is not biconnected. {\bf Q.E.D.}\medskip

\numpara
\label{witnesses} 
{\bf Witnesses for a non-nodally 3-connected graph.}
Suppose $G$ is not nodally 3-connected.  We
say that $H,K,u,v$ are {\em witnesses} if $G=H\cup K$,
$H\cap K$ contains just two vertices $u,v$ and no edge,
neither $H$ nor $K$ are path graphs, and neither $H$ nor $K$
equals $G$.

\begin{lemma}
\label{edges incident from H and K} 
{\rm (i)}
Given witnesses $H,K,u,v$, if $L$ is a path in $G$
connecting $H\backslash K$ to $K\backslash H$,
then $L$ contains three consecutive vertices $r,s,t$ where
$\{r,s\}\in H$, and $\{s,t\}\in K$, $r\in H\backslash K$,
$t\in K\backslash H$, and $s \in H\cap K$, so $s=u$ or $s=v$.

{\rm (ii)} Any path (respectively, cycle) which avoids $u$ and $v$ except
perhaps at its endpoints
(respectively, perhaps once), is entirely in $H$
or in $K$.
\end{lemma}

{\bf Proof.} (i) The first vertex in $L$ is in $H\backslash K$,
so the first edge is in $H$.  Similarly the last edge is in $K$.
Therefore there exist three consecutive vertices $r,s,t$
on the path where $\{r,s\} \in H$ and $\{s,t\}\in K$.
Then $s\in H\cap K$, so $s=u$ or $s=v$ and $s$ is
incident to edges from $H$ and from $K$.

(ii)
Now let $P$ be a path which avoids $u$ and $v$
except perhaps at its endpoints. This includes the
possibility of a cycle, viewed as a path which
begins and ends at the same vertex $w$: we allow
$w$, but no other vertex on the cycle, to equal
$u$ or $v$.

If the path is not entirely
in $H$ nor in $K$, then it contains a triple
$r,s,t$ where $s = u$ or $s=v$, a contradiction. {\bf Q.E.D.}\medskip

%

The proof of
Theorem \ref{planar nodally 3-connected} is long.  To lighten
it somewhat, we prove

\begin{lemma}
\label{2 or 3 faces} 
Let $G$ be a plane embedded graph in which all face
boundaries are simple cycles.  Then
{\rm (i)} either $G$ is a simple cycle with two faces, or\hfil\break
{\rm (ii)} for no two faces $F,F'$ is $\partial F \cap \partial F'$ a simple
cycle, and
if there are 3 faces $F_1,F_2,F_3$ such that
$$
Q_1 = \partial F_1 \cap \partial F_2,
Q_2 = \partial F_2 \cap \partial F_3, \quad\text{and}\quad
Q_3 = \partial F_3 \cap \partial F_1
$$ are all nonempty and connected, therefore simple paths, and they all
join the same two vertices $u$ and $v$,
then there are exactly three faces, and
$G$ consists of two nodes connected by three paths.
\end{lemma}

{\bf Proof.} Since all face boundaries are simple cycles,
$G$ is biconnected, hence connected.

(i) Suppose
$\partial F \cap \partial F' = \partial F,$  that
is $\partial F \cap \partial F'$ is a Jordan curve $J$.
By Theorem \ref{jordan curve theorem} (ii),
$F$ is the inside of $J$ and $F'$ the outside
or vice-versa, so $G$ is a simple cycle with two faces.

(ii)  W.l.o.g.\  $F_1$ and $F_2$ are bounded.  Their
intersection $Q_1$ is a simple path, which means that
$X=\overline{F_1}\cup \overline{F_2}$ is simply connected,
and $\partial X = \partial F_1 \cup \partial F_2 \backslash
\int( Q_1)$.

The only faces meeting the relative interior of $Q_1$ (respectively, $Q_3$) are
$F_1$ and $F_2$ (respectively, $F_3$ and $F_1$), so $Q_1 \not= Q_3$.
These are different paths joining $u$ to $v$ on $\partial F_1$,
so $\partial F_1 = Q_1\cup Q_3$. Again, $\partial F_2 = Q_1 \cup Q_2$,
Thus $\partial X = Q_2 \cup Q_3 = \partial F_3$.

$F_3$ is either the inside or outside of $\partial F_3$
(Theorem \ref{jordan curve theorem}), but $F_1\cup F_2$
are inside, so it is the outside, and $F_3$ is the
unbounded face.  Thus there are three faces and
$G$ is the union of three paths $Q_1\cup Q_2\cup Q_3$
with two nodes in common.
{\bf Q.E.D.}\medskip

\begin{theorem}
\label{planar nodally 3-connected} 
A plane (straight-edge)
embedded graph is nodally 3-connected iff it is biconnected and
the intersection of any two face boundaries is connected.
\end{theorem}

{\bf Proof.} We can assume $G$ is biconnected,
since that is required for nodal 3-connectivity.
Since $G$ is biconnected either it is empty or
trivial, or a single edge, or
every face is bounded by a simple cycle.
In the first three cases the graph is obviously
nodally 3-connected and
biconnected with one face, so we need
only consider the fourth case and can assume that every face is bounded by
a simple cycle.

We can assume that $G$ is straight-edge embedded. Therefore
the boundary of every face is a simple polygon.

{\bf Only if:} Suppose $F_1$ and $F_2$ are different faces and
$\partial F_1\cap \partial F_2$ is disconnected. R.T.P.
$G$ is not nodally 3-connected.

Let
$u$ and $v$ be vertices in different components of
$\partial F_1\cap \partial F_2$.  For $i=1,2$
there are two paths $P_i$ and $Q_i$ joining
$u$ to $v$ in $\partial F_i$.  These paths
are polygonal.

One can also construct a path $P_1'$ within $F_1$, loosely speaking
by displacing $P_1$ slightly into $F_1$, and connecting its endpoints
to $u$ and $v$.  The resulting path is in
$F_1$ except at its endpoints.  Similarly one can construct
a path $P_2'$ in $F_2$ except at its endpoints.
These paths together form a (polygonal) Jordan curve $J$ which 
meets $G$ only at $u$ and $v$.  By construction, $P_1 \cup P_2$
is inside $J$ and $Q_1 \cup Q_2$ is outside $J$.

Let $H$ (respectively, $K$) be the subgraph consisting of all
vertices and edges of $G$
which lie inside or on $J$ (respectively, outside or on $J$).
The only vertices in $H\cap K$ are $u$ and $v$, and $H\cap K$
contains no edge. $H$ contains
$P_1 \cup P_2$ and therefore is not a path graph, since otherwise
$P_1 = P_2$ and $u$ and $v$ would be in the same component of
$\partial F_1 \cap \partial F_2$.
Similarly $K$ is not a path graph.  Therefore
$G$ is not nodally 3-connected.\hfil\break

{\bf If:}  Suppose $G$ is biconnected but not nodally 3-connected,
and $H,K,u,v$ are witnesses.  $G$ has more than one face,
so all face boundaries are simple cycles.

{\em Claim 1.} The subgraphs
$H\backslash K$ and $K\backslash H$ are nonempty.
If every vertex in $K$ were also in $H$,
then the vertices in $K$ are in $H\cap K$,
that is, $u$ and $v$.  Either $K$ has no
edges, in which case $H=G$, or
it has the edge $\{u,v\}$ and is a path graph.
Neither is possible.  Therefore $H\backslash K$
and similarly $K\backslash H$ are nonempty.

{\em Claim 2.} Neither $u$ nor $v$ are isolated
vertices in $H$ nor in $K$.

Otherwise suppose $u$ is isolated in $K$.
Let $L$ be any path joining $H\backslash K$ to $K\backslash H$.
By Lemma \ref{edges incident from H and K}, every
path connecting $H\backslash K$ to $K\backslash H$
contains a vertex, $u$ or $v$, incident to
edges from $H$ and from $K$.  By hypothesis, $u$
is not; so every such path contains $v$.  By
Claim 1, at least one such path exists, so
$G\backslash v$ is not connected, and $G$
is not biconnected.

{\em Claim 3.}
Both $u$ and $v$ have neighbours both in $H\backslash K$ and
in $K\backslash H$. Suppose all neighbours of $u$ are in $H$.
Since $u$ is not isolated in $K$, there is an edge
$\{u,t\}$ in $K$ incident to $u$.  But $t$ is a neighbour
of $u$, therefore $t\in H\cap K$, so $t=v$.  The only
edge in $K$ incident to $u$ is $\{u,v\}$.

Consider a path in $G$
joining $H\backslash K$ to $K\backslash H$. Let
$t$ be the first vertex where the path meets $K\backslash H$,
and let $s$ be the vertex before $t$ on the path. Since
$\{s,t\}\in K$ and $s\notin K\backslash H$, $s\in H\cap K$:
$s=u$ or $s=v$.  However, if $s=u$, then, since $t\in K$,
$t=v$ and $t\notin K\backslash H$.  Therefore $s=v$.  This
implies that every path from $H\backslash K$ to $K\backslash H$
contains $v$. Again by Claim 1, such paths exist, so
$G$ is not biconnected.

This contradiction shows
that not all neighbours of $u$ are in $H$; neither are
they in $K$, and the same goes for $v$.

{\em Claim 4.} The vertices $u$ and $v$ share a face in common.
Otherwise let $x_1,\ldots,x_k$ be the neighbours
of $u$.  We know (Lemma \ref{link of u}) that
they are all connected by paths in $B\backslash u$,
where $B$ is the union of boundaries of bounded
faces incident to $u$.  Assuming $v$ is incident to none
of these faces, these
paths would also avoid $v$. This implies that all neighbours
of $u$ are in $H$ or in $K$, contradicting Claim 3.

{\em Claim 5.} The vertices $u$ and $v$ have at least two
faces in common.  Let
$F_1,\ldots$ be the faces incident
to $u$ in anticlockwise order around $u$.  At least
one of these faces, w.l.o.g.\ $F_1$, is incident to $u$ and to $v$.
Suppose no other face is.

There are two cases.
If $u$ or $v$, w.l.o.g.\ $u$, is an internal vertex, then
all faces incident to $u$ are bounded, and by Lemma
\ref{link of u}, the subgraph
$\bigcup_{i\geq 2} (\partial F_i \backslash u)$ would
be connected and contain neither $u$ nor $v$.  Then
all vertices in this subgraph would belong to $H$
or to $K$.  Since it includes all neighbours of $u$
in $G$, it would contradict Claim 3.

If both $u$ and $v$ are external vertices, then $F_1$
is the external face,
and all bounded faces incident to $u$ avoid $v$.
This time we consider the subgraph
$\bigcup_{i\geq 2} (\partial F_i \backslash u)$.
Again this is
a connected subgraph containing all neighbours of $u$
in $G$, and again it omits both $u$ and $v$, so
again all vertices in it are in $H$ or in $K$, and
again Claim 3 is contradicted.

Therefore $u$ and $v$ have at least two faces $F$ and $F'$ in common.

{\em Claim 6.}
If $u$ and $v$ are incident to three faces
$F_1$, $F_2$, and $F_3$, then the boundaries of at least two
of these faces have disconnected intersection.
Otherwise,
by Lemma \ref{2 or 3 faces}, $G$ consists of two
nodes $u,v$ connected by three paths. If $G= H\cup K$
where $H\cap K = \{u,v\}$ then $H$ or $K$ is
a path graph: $G$ is nodally 3-connected.

This contradiction shows that the one of the pairs
$\partial F_i \cap \partial F_j$ is disconnected, as
claimed.

{\em Claim 7.}
If there are exactly two faces $F$ and $F'$ incident
to $u$ and to $v$,
then $\partial F\cap \partial F'$ is disconnected.

Otherwise $\partial F\cap \partial F'$
is a path $Q'$ joining a vertex $u'$ to another
vertex $v'$ and containing a
subpath $Q$ joining $u$ to $v$.  Not all of
$u',u,v,v'$ need be distinct, but it is assumed
that they occur in that order in $Q'$.

By
Lemma \ref{edges incident from H and K},
all vertices in $Q$
belong to $H$ or to $K$: w.l.o.g.\ 
to $H$. The boundary cycles  $\partial F$ and
$\partial F'$ include two other paths, $Q_1$ and $Q_2$,
respectively, joining $u'$
to $v'$. Let $J=Q_1\cup Q_2$, a Jordan curve.

If $u'\not= u$ then $J$ meets $H\cap K$ at $v$ alone,
or not at all, and by Lemma
\ref{edges incident from H and K}, all vertices on
$J$, plus those in $Q'\backslash Q$, belong to $H$ or to $K$.

If all vertices on $J$ belong to $H$, then all
vertices outside $J$ also belong to $H$, because for
any vertex $y$ outside $J$, one can choose a shortest
path joining $y$ to a vertex in $J$.
Neither $u$ nor $v$ occur as internal vertices on this
path, so all vertices on the path are in $H$ or $K$ (Lemma
\ref{edges incident from H and K}),
i.e., $H$, since the last vertex is in $H$.

We have counted all vertices in $G$: those outside $J$,
those on $J$, and those on $Q'$, and all are in $H$,
so $H=G$, which is false.

On the other hand,
if all vertices on $J$, and in $Q'\backslash Q$, belong to $K$,
then all vertices outside $J$ belong to $K$,
and $H=Q$ is a path graph, which is false.
This proves Claim 7 in the case $u\not= u'$,
and by symmetry in the case $v\not= v'$.

If $u=u'$ and $v=v'$ then $Q=Q'$: let $Q_1$ and
$Q_2$ be the other subpaths joining $u$ to $v$
in $\partial F$ and $\partial F'$ respectively.
By Lemma \ref{edges incident from H and K},
each subpath $Q_i$ is contained in $H$ or
in $K$.  Again we have a Jordan curve
$J = Q_1 \cup Q_2$.

If $u$ and $v$ are not both external vertices,
w.l.o.g.\
$u$ is an internal vertex, then
$F$ and $F'$ are bounded faces incident to $u$,
and since $\partial F \cap \partial F' = Q$,
they are consecutive in cyclic order.
Let $u_1$ (respectively, $u_2$) be the second
vertex (following $u$) in $Q_1$ (respectively, $Q_2$).
The only faces incident to $u$ and to $v$ are 
$F$ and $F',$ so $u_1$ and $u_2$
differ from $v$ and
$u_1$ and $u_2$ are connected
by a path which avoids $u$ and $v$
(Lemma \ref{link of u}).
Therefore, by Lemma 
\ref{edges incident from H and K}, $u_1$ and $u_2$ are both
in $H$ or in $K$, and so are all vertices on $J$.
The same goes
for all vertices outside $J$, so either $H=G$ or
$H=Q$ is a path graph, a contradiction.

This leaves the case where $u$ and $v$ are external
vertices with exactly
two faces in common, $F$ and $F'$, whose
boundaries have connected intersection.
Since $u$ and $v$ are external vertices,
one of these faces,
$F',$ say, is the external face.
Since $G$ is not nodally 3-connected, it is not
a simple cycle, and $Q = \partial F\cap \partial F'$ is a simple path
joining $u$ to $v$ (Lemma \ref{2 or 3 faces}).
Let $Q_1$ and $Q_2$ be the other paths
joining $u$ to $v$ on $\partial F$ (respectively,
$\partial F'$). $\partial F' = Q \cup Q_2$ is the
external cycle, a Jordan curve, and $Q_1$ separates
its interior into two regions of which $F$ is one.
Let $J=Q_1\cup Q_2$. It is a Jordan curve surrounding
the other region.

Let $u_i,~ i=1,2,$ be the second vertices
on $Q_i$. Again
there is a path joining $u_1$ to $u_2$ which
avoids $u$ and $v$, and
all vertices on $J$ are in $H$ or $K$, and
the same holds for all vertices inside $J$.
If they are all in $H$ then $H=G$, and if they
are all in $K$ then $H=Q$, a simple path.
This contradiction finishes the proof of Claim 7.

Claims 6 and 7 taken together amount to the desired result.
{\bf Q.E.D.}\medskip

\numpara
\label{chord-free triangulated graphs} 
{\bf Chord-free triangulated graphs.}
A triangulated plane embedded graph is one in which every
bounded face is bounded by three edges.  In a triangulated
biconnected graph the external boundary is also a simple cycle.
It can only fail to be nodally 3-connected if a bounded
face meets the external boundary in a disconnected set.
Equivalently, one of its edges is a chord joining
two vertices on the external boundary,
and the other two edges are not both on the external boundary
\cite{white}.

\begin{figure}
\centerline{\includegraphics[height=1in]{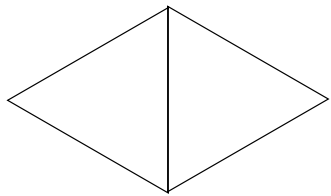}}
\caption{a nodally 3-connected but not triconnected triangulated
planar graph}
\label{quadril.fig}
\end{figure}

The graph in Figure \ref{quadril.fig} is nodally 3-connected but
not triconnected.

A fully triangulated planar graph is a triangulated planar
graph in which there are three external edges. In other words,
the external face also is bounded by a 3-cycle.  Therefore the
external cycle has no chords, so every fully triangulated
planar graph is nodally 3-connected.

Also let $G$ be a fully triangulated planar graph containing
a vertex $v$ of degree 2. Let $u$ and $w$
be the neighbours of $v$. There are only two faces
incident to $v$ and they are both incident to
$u,v,$ and $w$.  One of them must be
the external face. Thus $u,v,$ and $w$ are the
three external vertices.  They also bound
the only bounded face.  $G$ is a 3-cycle,
and therefore triconnected.

On the other hand,
if $G$ is fully triangulated
then it is nodally 3-connected,
so if it contains no vertex of degree $2$
then it is triconnected
(Proposition
\ref{nodally 3-connected and no deg 2}).  Therefore

\begin{corollary}
Every fully triangulated planar graph is triconnected.\qed
\end{corollary}

\section{Conditions for a convex combination map to be an embedding}
\label{convex combination embedding section} 
In this section we consider a plane embedded graph $G$ whose
external boundary is a simple cycle.

\numpara
\label{par: embedding means convex embeddable}
If a convex combination map of $G$ is an embedding,
then it is a convex embedding
(Lemma \ref{lem: conv comb emb conv faces}),
so every face boundary
is a simple cycle and the intersection of every two bounded faces is
convex, hence connected.  Also,
if a bounded face meets both ends of an external
edge, then it is incident to that edge. This
gives three conditions necessary for the
existence of a convex embedding, and hence for
a convex combination map.

The first two conditions, and a weakened version of the third,
were given by Stein \cite{stein}, investigating the
existence of convex embeddings.
He allowed new vertices to be added within edges
so effectively edges are mapped to polygonal curves,
weakening the third condition in the following
definition.

\begin{figure}
\begin{center}
\includegraphics[width=1.5in]{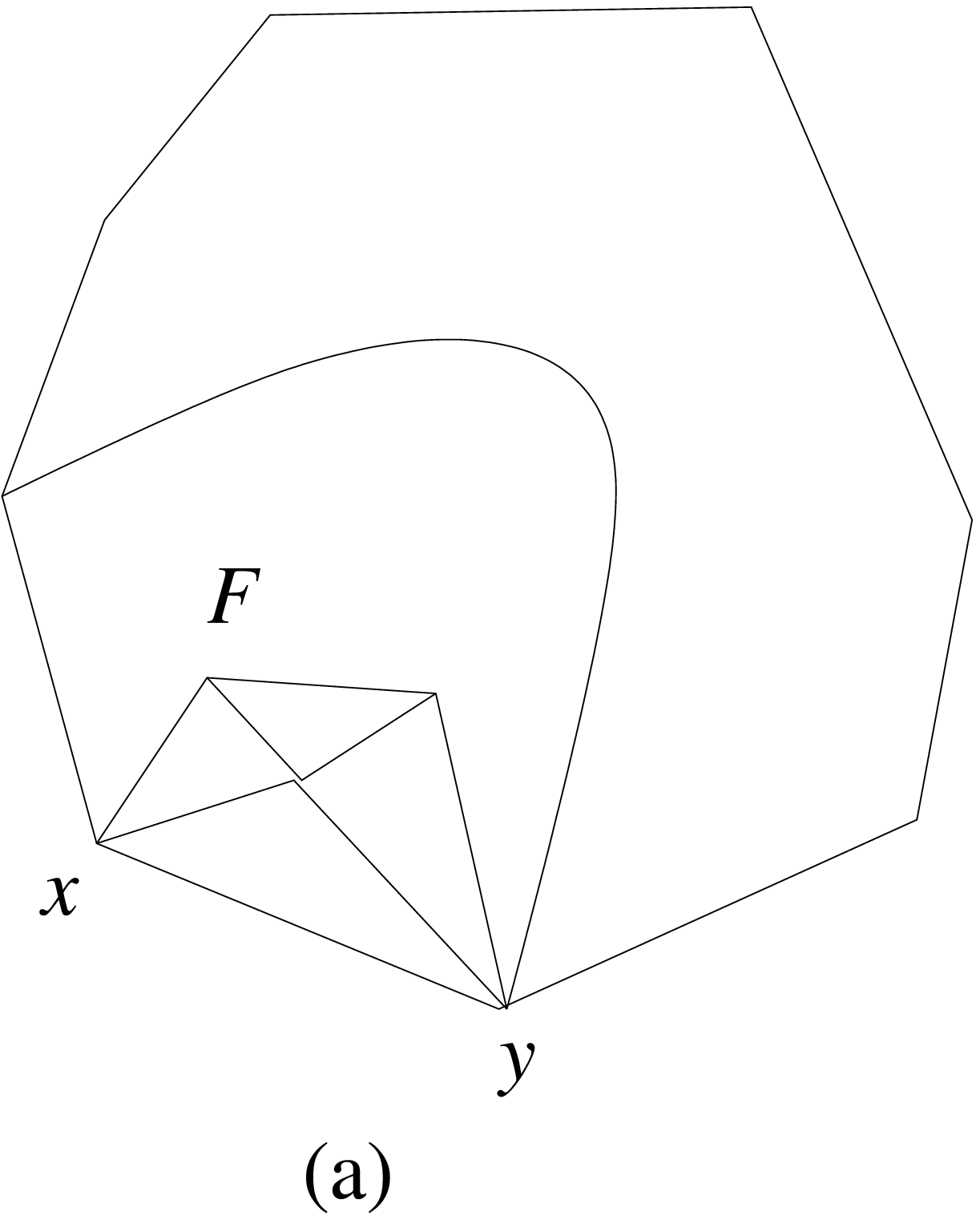}
\hfil
\includegraphics[width=1in]{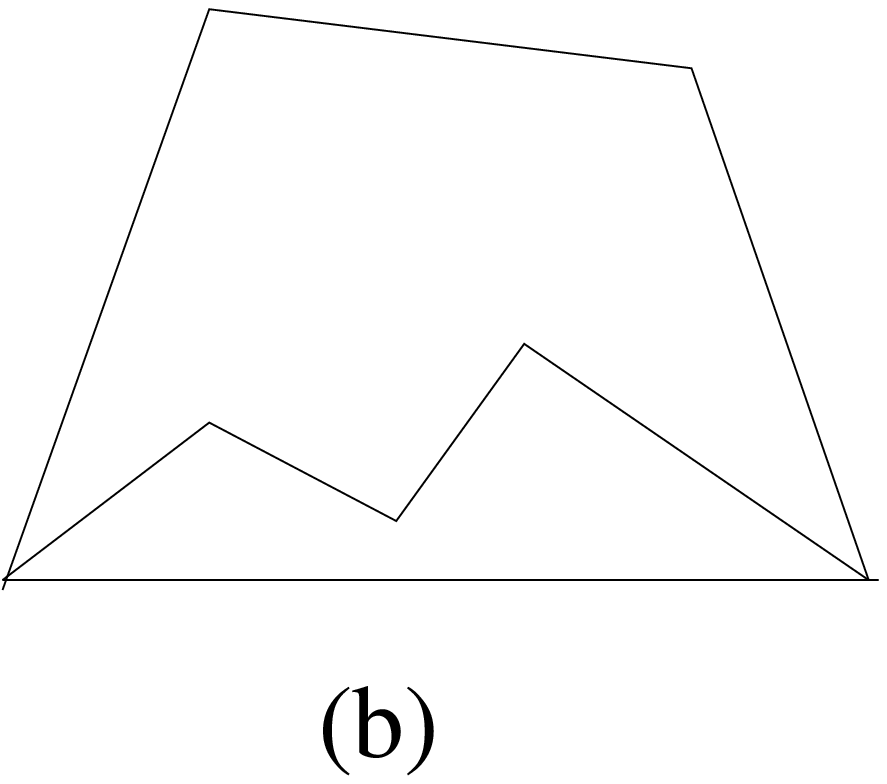}\\
\caption{(a) inverted subgraph. (b)
A nodally 3-connected graph which is not convex embeddable.}
\label{inverted.fig}
\label{nontutte2.fig}
\end{center}
\end{figure}

\begin{definition}
\label{inverted subgraphs} 
\label{convex embeddable} 
Let $G$ be a plane embedded biconnected graph.
If a bounded face $F$ meets both ends of
an external edge, but $F$ is not incident to
that edge, then the subgraph between $F$
and that edge is called an {\em inverted subgraph.}
See Figure \ref{inverted.fig}.

$G$ is {\em convex embeddable} if $G$ is nonempty,
every face boundary is a simple
cycle, the intersection of every two bounded faces is
connected, and there are no inverted subgraphs.
\end{definition}

The phrase `convex embeddable' suggests that $G$ admits
a convex embedding, and this will prove to be true (Theorem
\ref{embedding theorem}).
The phrase `inverted subgraph' is used because it is
possible, by repeatedly reflecting inverted subgraphs
through the external boundary, to produce an embedding
in which there are no inverted subgraphs.

\begin{lemma}
\label{lem: embedding means convex embeddable}
If some convex combination map of $G$ is an embedding
then $G$ is convex embeddable (immediate from
Paragraph \ref{par: embedding means convex embeddable}).\qed
\end{lemma}

There is one class of nodally 3-connected plane-embedded graphs
which are not convex embeddable.  These graphs have two
nodes joined by three paths of which one is
an external edge (Figure \ref{nontutte2.fig}).
Apart from these graphs, every nodally 3-connected
graph is convex embeddable.

The aim of this section is to prove  that if $G$ is
convex embeddable, then every convex combination map of $G$
is an embedding.  This has already been shown
by Floater for triangulated planar graphs
(\ref{floaters theorem}) and we shall depend heavily
on that result.
The point here is that we can consider limiting cases of
convex combination maps, which would make no
sense for barycentric maps.
Rather than taking the more obvious approach
and attempting induction on the number of faces of $G$,
we can use Floater's result to describe a convex
combination map $f$ as a {\em limit} of straight-edge
embeddings $f^\delta$.

For the remainder of the section, $G$ will be a
plane embedded graph whose boundary is
a simple cycle, and $f$ a convex combination
map of $G$.  We shall use $P$ to denote the convex polygon
whose corners are the images of external vertices,
Also, $\lambda_{uv}$ are the coefficients associated
with $f.$

\begin{lemma}
\label{injective on external vertices} 
If $G$ is biconnected, and $u$
is an external vertex, then
for all vertices $v\not= u$, $f(v) \not= f(u)$.
\end{lemma}

{\bf Proof.} 
Since $u$ is external, $f(u)$ is a corner of $P$,
and it is not a proper convex combination
of any other subset of $\hull(P)$.

Let $S$ be the set of all vertices $v$ such that
$f(v) = f(u)$.  Note that $u$ is the only external
vertex in $S$.

We assume that $S$ contains some vertex besides $u$,
or equivalently, $S$ contains at least one internal vertex.

No internal vertex $v\in S$
can be adjacent to any vertex $w\notin S$.
Otherwise $v$ would have a neighbour $w$ with $f(w) \not= f(v)$,
$f(v)$ would be a proper convex combination of points in $\hull(P)$
including $f(w)\not= f(v) \in \hull(P)$,
and $f(v)$ would not be a corner of $\hull(P)$.

Let $H$ be the subgraph of $G$ spanned by $S$:
$$ H = ( S, \{ \{u,v\} \in G:~ u,v \in S\}).$$ Claim
that $H$ is connected. Otherwise it has
a connected component $K$ not containing $u.$
All
vertices in $K$ are internal vertices of $G$,
so all vertices adjacent (in $G$) to vertices
in $K$ are also in $K$: $K$ is a connected component of $G$
not containing $u$, so $G$ is disconnected, proving the claim.

Since $H$ contains other vertices besides $u$, and is connected,
it contains an internal vertex $v$ adjacent to $u.$

Therefore $u$ is adjacent in $G$ to a vertex $v \in S$.
Since $u$ is adjacent to two external vertices,
$u$ is also adjacent to a vertex $w \notin S$.

Let $\Pi$ be a path from $v$ to $w$ in $G.$ There must be
at least two consecutive vertices $x,y$ in $\Pi$
where $x\in S$ and $y\notin S$. Then $x$ cannot be an internal node,
so $x = u$.
This shows that every
path from $v$ to $w$ contains $u$, so
$G\backslash\{u\}$ is disconnected and $G$ is
not biconnected. {\bf Q.E.D.}\medskip

\numpara
\label{par: definition of the maps f delta}
{\bf Definition of the maps $f^\delta$.}
Let $G'$ be obtained by triangulating $G$
(Proposition \ref{all embeddings triangulable}).
$G$ and $G'$ have the same vertices, the
same internal vertices, and the same external
vertices.  For each internal vertex $u$ let
$\Gamma_u$ be its neighbours in $G$ and
$\Gamma'_u\supseteq \Gamma_u$ its neighbours in $G'.$

For
any $\delta$, $0\leq\delta<1$, internal vertex $u$ and vertex $v$,
let
$$ \lambda^\delta_{uv} = \lambda_{uv}
\quad\text{if}~ \Gamma'_u \backslash \Gamma_u = \emptyset.$$ Otherwise,
$\Gamma'_u\not= \Gamma_u$:
$$
\lambda^\delta_{uv} = \begin{cases}
\frac{\delta}{| \Gamma'_u \backslash \Gamma_u|}
\quad\text{if}~v\in \Gamma'_u \backslash \Gamma_u,\\
(1-\delta)\lambda_{uv}
\quad\text{if}~v\in \Gamma_u ,\\
0\quad\text{otherwise.}
\end{cases}
$$ If $u$ is an external vertex,
$\lambda^\delta_{uv} = 1$ if $u=v$ and $0$ if $u\not=v$,
just as with $\lambda_{uv}$. (See Definition \ref{convex combination map}.)

\begin{definition}
\label{f sup delta} 
With $G,G',f$ and $\delta$ as just introduced,
let $f'$ be the convex combination map of $G'$ with
with coefficients $\lambda^\delta_{uv}$ and
$f'(x) = f(x)$ for each external vertex $x$.
This is a straight-edge
embedding if $\delta > 0$ (Proposition \ref{floaters theorem}).

We define $f^\delta$ as the restriction of $f'$ to $G$.
\end{definition}

Recall (Paragraph \ref{matrix defining a convex combination map})
that $f$ and $f^\delta$ can be identified with column
vectors of height $2m$, which allows us to define
the distance between them.  It is most natural to
define
$$ || f - f^\delta || =
\max \{ |f(v)-f^\delta(v)|:~ v~\text{a vertex}\}.
$$

\begin{lemma}
\label{f is limit of f delta} 
$\lim_{\delta\to 0} f^\delta = f.$
\end{lemma}

{\bf Proof.} The map $f$ is the unique solution to
$AX=B$, and
$f^\delta$ is the unique solution to
equations of the form $(A+\delta A')X = B$,
where $A$ is the matrix defining $f$. Also $A$
is invertible
(Lemma \ref{unique convex combination map}), so
for small $\delta$ the map $\delta \mapsto (A+\delta A')^{-1}$
is well-defined and continuous. Therefore, as $\delta\to 0$,
$f^\delta \to f$. {\bf Q.E.D.}\medskip

\numpara
\label{remarks about f sup delta} 
{\bf Remarks about the map $f^\delta$.}

\begin{itemize}
\item
The map $f^\delta$ is a straight-edge
embedding if $\delta>0$, but
is the restriction of a convex combination map $f'$
of a triangulated graph,
not itself a convex combination map of $G$.
Face boundaries are mapped to
simple polygons under $f^\delta$.  They are not necessarily convex.
\item
$f^0 = f$ is a convex combination map of $G$.
\item
Since $f = f^0 = \lim_{\delta\to 0} f^\delta$,
even though $f$ might not be an
embedding, it fails to be only because
edges may collapse to points and faces
collapse to line-segments or points.
\item
The map
$f$ partially preserves the cyclic order of
edges around a vertex, but edges may collapse to points
or consecutive
edges may overlap.
The interpretation is that the face between them
has collapsed under $f$.
\end{itemize}

%
%
%
%

\numpara
\label{homeomorphism from f delta} 
{\bf Extending $f^\delta$ to a homeomorphism.}
The graph $G'$
is a plane embedded graph and all its bounded faces
are bounded by 3-edge Jordan curves. It can
be arranged that $G'$ is embedded with
straight edges, hence so is $G$. Fix $\delta,~
0 < \delta < 1.$  Let $f'$ and $f^\delta$ be defined as above.

By Floater's result
(Proposition \ref{floaters theorem}),
$f'$ is a straight-edge embedding of $G'$.
Let $u,v,w$ be the three vertices on the boundary of a
bounded (triangular)
face of $G'$.  The map $f'$ can be extended in a piecewise-linear
fashion to this face and all bounded faces.
Let $\overline{G}$ be the complement of the
unbounded face of $G$ (and of $G'$).
The map $f'$ extended to the bounded faces of $G'$ is
a piecewise-linear homeomorphism from $\overline{G}$ onto
$\hull(P)$.
This homeomorphism can also be written
as $f^\delta$.

Thus  $f^\delta$ means either a
straight-edge embedding of $G$ or
a piecewise-linear homeomorphism from
$\overline{G}$ onto $\hull(P)$.

\begin{definition}
\label{def: degenerate}
An edge $e$ is {\em degenerate} if $f(e)$ is
a single point.
\end{definition}

\begin{lemma}
\label{lem: vertex not interior edge}
For any nondegenerate edges $e_1$ and $e_2$,
$f(e_1)$ does not meet the interior of $f(e_2)$ transversally.
\end{lemma}

{\bf Proof.} Suppose otherwise.  The interiors of $f(e_1)$
and $f(e_2)$ cannot intersect transversally, since
otherwise for some $\delta>0$ the interiors of $f^\delta(e_1)$ and
$f^\delta(e_2)$ would intersect transversally.  Suppose that $e_1 = \{u,v\}$
and $f(v)$ is interior to $f(e_2)$.  Let $L$ be the line
through $f(e_2)$.  The vertex $u$ is a neighbour of $v$ such
that $f(u)\notin L$, and $v$ cannot have another neighbour $w$
such that $f(w)$ is on the other side of $L$, since otherwise
for some $\delta>0$ the line segment $f^\delta(e_2)$  and
the broken line $f^\delta(u)f^\delta(v)f^\delta(w)$ would 
intersect in their interiors.
Also, $f(v)$ is interior to $f(e_2)$, hence inside $P$,
and $v$ is an internal vertex.  This contradicts Lemma
\ref{lem: one-sided}.
{\bf Q.E.D.}\medskip

The following proposition is a simple corollary to
the Jordan Curve Theorem.

\begin{proposition}
\label{prop: interlacing}
{\bf (interlacing property).}
Let $J$ be a Jordan curve and
$a,b,c,d\in J$ be four points in cyclic order around $J$.
If $X$ and $Y$ are paths inside
$J$ meeting $J$ only at $a$ and $c$, $b$ and $d$, respectively,
then $X$ and $Y$ intersect inside $J$. \qed
\end{proposition}

\begin{lemma}
\label{lem: face 3 edges then collinear}
If $F$ is a (bounded) face where $\partial F$ is a simple cycle,
and $p$ is a point such
that for three or more edges $e $ on $\partial F$,
$f(e)$ is nondegenerate and incident to $p$,
then all edge-images $f(e)$, which are
incident to $p$, are collinear.
\end{lemma}

{\bf Proof.}
Let $\partial F= v_1,\ldots,v_n$,
\be
\eta\quad=\quad
\frac{ \min \{ |f(v)-p|: ~ v~\text{a vertex and}~ f(v)\not= p\}}{2},
\ee
and $D$ be the closed disc
with centre $p$ and radius $\eta$.

For every vertex $v$, $f(v)\in D\iff f(v)=p$.
Choose $\varepsilon > 0$ so that for
all $\delta$ with $0\leq \delta \leq \varepsilon$
and every vertex $v$, $f^\delta(v)\in D \iff f(v)=p$.

Given adjacent vertices $u$ and $v$ on $\partial F$
such that $f(v)=p$ and $f(u)\not= p$, suppose $u= v_{i_1}$.
Beginning with
$u,v\ldots$, traverse $\partial F$ in cyclic order 
until the next vertex $v_{i_2}$ is reached such that
$f(v_{i_2})\not= p$.  Continue the traversal in cyclic
order until the next such pair $u,v$ is found,
hence identifying a subpath $v_{i_3},\ldots,v_{i_4}$, and
continue in this way until $\partial F$ has
been traversed fully.
In this
way we get a series $I_1 = v_{i_1},\ldots, v_{i_2}$, $I_2$,\ldots
$I_k$,
of paths in $\partial F$, joining vertices
$v_{i_j}$ to $v_{i_{j+1}}$ ($j=1,3,5\ldots$)
where $f(v_{i_j})\notin D$ for all $j$, and all inner
vertices (Paragraph \ref{subgraphs etcetera})
 in each path $I_j$ are mapped to $p$. By hypothesis,
$k \geq 2$.

For $1\leq j \leq k$
let $V_j$ be the set of inner vertices in $I_j$,
and let $U_j$ consist
of every vertex in $G$ which is not in $V_j$
but which has a neighbour in $V_j$.  $U_j$ can
include vertices not in $\partial F$.

Since every two vertices in $V_j$ are connected
by a path in $V_j$, every two vertices $a,b$ in $U_j$
are connected by a (unique) simple path $P_{ab}$ 
whose inner vertices are
in $V_j$.  The image $f(P_{ab})$
is the polygonal path $f(a) p f(b)$.

{\bf Claim:} given $a,b  \in U_1$ and $c,d\in U_2$,
the paths $f(a)pf(b)$ and $f(c)pf(d)$ do not
cross, meaning that given $\partial D\cap pf(a)=a'$,
with $b',c',d'$ similarly defined, the points

\centerline{
$ a',c',b',d' $}

\noindent are {\em not} in strict cyclic
order around $\partial D$.

Otherwise
let $X=D\cap f^\varepsilon(P_{ab})$ and
$Y=D\cap f^\varepsilon(P_{cd})$.
The endpoints of $X$ and $Y$ are alternating in cyclic
order around
$\partial D$. By Proposition
\ref{prop: interlacing},
$X$ and $Y$ intersect
in the interior of $D$. Since $f^\varepsilon$
is an embedding, the intersection
is contained in  $f^\delta(V_1\cap V_2)$, whereas
$V_1\cap V_2 = \emptyset$.  This contradiction proves the claim.

Let $C_j$ be the set of points on $\partial D$
where edge-images $f(u)f(v)$, $u\in U_j, v\in V_j$,
intersect $\partial D$.  Let $c_j$ be the
smallest arc of $\partial D$ containing
$C_j$.  This is ambiguous only when $k=2$
and $C_1=C_2$ contains two diametrically
opposed points, in which case we may choose
$c_1$ and $c_2$ either way (but different).

By the above claim,
$c_1$ and $c_2$ do not overlap.  Hence they
cannot both subtend reflex angles at $p$.
Without loss of generality, $c_1$ subtends
an angle $\alpha\leq 180^\circ$ at $p$.  Let $L$
be a line through $p$ which does not intersect
the relative interior of $c_1$.  Then for all neighbours
$u$  of $v_{i_1+1}$ in $G$, $f(u)$ is on $L$, or on
the same side of $L$ as is $f(v_{i_1})$.  But
since there exists more than one vertex $v$
such that $p=f(v)$, $v_{i_1+1}$ is an internal
vertex  (Lemma \ref{injective on external vertices}),
and $f(v_{i_1+1})$ is a proper weighted average of its neighbours.
By Lemma \ref{lem: one-sided},
$C_1 = \partial D \cap L$ and $\alpha=180^\circ$.
Therefore $c_2$ does not subtend a
reflex angle at $p$, and by the same argument
$C_2 = \partial D \cap L$.  Therefore $c_1 \cup c_2 = \partial D$,
$k=2$, and for all
edges $\{u,v\}$ with $f(v) = p$, $f(u)\in L$,
as claimed. {\bf Q.E.D.}\medskip

As already mentioned, this section aims to
prove that if $G$ is convex embeddable
then $f$ is an embedding.
We show that there are no degenerate edges, and
therefore $f$ is injective on faces.
It will follow by Tutte's argument \cite{tutte} that $f$
is an embedding.
We first study what happens
if $f$ collapses faces, and this leads us to consider
the notion of monotone paths.
The definition needs to allow for the possibility
that $f$ maps different vertices to the same point.

\begin{definition}
\label{def: monotone path}
Given $0\leq \varepsilon < 1$ and a line $V$,
$V$ is {\em $\varepsilon$-vertex-avoiding}
or simply {\em vertex-avoiding} when $\varepsilon = 0$,
if, for all $\delta \leq \varepsilon$,
and all vertices $v$, $f^\delta(v)\notin V$.

Let $V$ be a directed vertex-avoiding line. Given
nondegenerate edges $e_1,e_2$, $e_1$
is {\em above} $e_2$ on $V$ if $V$ intersects
the relative interiors of $f(e_i)$, $i=1,2$, and for some
$\varepsilon$ such that $V$ is $\varepsilon$-vertex-avoiding,
$V$ intersects
the relative interiors of $f^\varepsilon (e_i)$ at points
$a_i$ where $V$ is directed from $a_2$ to $a_1$.

Let $L$ be a directed line.
A path $v_i,\ldots, v_k$ in $G$ is {\em monotone} (on $L$)
if all points $f(v_i),\ldots, f(v_k)$ belong to $L$ and are monotone
non-decreasing or monotone non-increasing on $L$.

Given two paths $s_1$ and $s_2$ which are monotone on $L$, and
which have no vertices in common except perhaps at endpoints, we say that
$s_1$ is {\em above} $s_2$ if there exists a directed vertex-avoiding line
$V$ positively normal to $L$ and edges $e_i\in s_i$ such
that $e_1$ is above $e_2$ on $V$.
(See \cite{edelsbrunner}, \S 11.2.)
\end{definition}

\begin{lemma}
\label{lem: asymmetric}
If $s_1$ and $s_2$ are monotone on $L$, and
they are vertex-disjoint except perhaps at endpoints,
and $s_1$ is above $s_2$, then $s_2$ is not above $s_1$.
\end{lemma}

{\bf Proof.} Given $V$ and edges $e_1$ on $s_1$ and $e_2$
on $s_2$, and $\varepsilon > 0$ so that $V$ is
$\varepsilon$-vertex-avoiding, then the relative order of
$V\cap f^\delta(e_1)$ and $V\cap f^\delta(e_2)$ is
unchanged for $0<\delta\leq \varepsilon$, since otherwise
for some $\delta>0$ $f^\delta(e_1)\cap f^\delta(e_2)\not= \emptyset$.
So if $f^\delta(e_1)$ is above $f^\delta(e_2)$ on $V$ for $\delta=\varepsilon$
then it holds for all positive $\delta \leq \varepsilon$.

Again, suppose that $s_2$ is also above $s_1$
according to different data $V',e_1',e_2',\varepsilon'$.
We can replace $\varepsilon$
and $\varepsilon'$ by their minimum and assume $\varepsilon=\varepsilon'$.
We could enclose these path-images by rectangles
bounded on two sides by $V$ and $V'$; the intersection points have the
interlacing property so $f^\varepsilon(s_1)$ and
$f^\varepsilon(s_2)$ would intersect in
their interiors
(Proposition \ref{prop: interlacing}), which is impossible.
{\bf Q.E.D.}\medskip

If $e\in \partial F$ and
$\partial F$ is a simple cycle (which is always true
when $G$ is biconnected), then for any $\varepsilon>0$,
$f^\varepsilon(F)$ is incident to $e$ from just one side.

\begin{lemma}
\label{lem: no flips}
Let $G$ be biconnected, $F$ a face, and $e=\{u,v\}$ a nondegenerate
edge in $\partial F$. Let $E$ be the directed line-segment
$f(u)f(v)$ and
for any $\varepsilon$, $0<\varepsilon <1$, let
$E^\varepsilon= f^\varepsilon(u)f^\varepsilon(v)$.
Then if $\varepsilon$ is sufficiently small, for
$0<\delta \leq \varepsilon$, $f^\delta(F)$ is always on
the same side (right or left) of $E^\delta$.
\end{lemma}

{\bf Proof.} Let $V$ be a vertex-avoiding line intersecting $E$,
and choose $\varepsilon>0$ so that $V$ is $\varepsilon$-vertex-avoiding.
Given $0 < \delta \leq \varepsilon$, let
$X^\delta=V\cap f^\delta(\partial F)$.
If $F$ is the external face then $f^\delta(\partial F)=P$
and the result is trivial.  We may assume that $F$ is
bounded so for all $\delta > 0$
$f^\delta(\partial F)$ is a simple polygon containing $f^\delta (F)$.

$X^\delta$
divides $V$ into open intervals
alternately inside and outside $f^\delta(F)$.
Also, $f(F)$ is to the right of $E^\delta$ if and only if
the number of points in $X^\delta$ to the left of $E^\delta$
is even. By choice of $\varepsilon$ this number is constant
for $0 < \delta \leq \varepsilon$. {\bf Q.E.D.}\medskip

\begin{definition}
\label{def: reflex corner}
Let
$F$ be a bounded face with $\partial F$ a simple
cycle $v_1,\ldots,v_n$: $f(\partial F)$ is a possibly
degenerate polygon, a union
of $k$ line-segments $p_i p_{i+1}$ (interpreting $p_{k+1}$
as $p_1$):
$p_1=f(v_1)$; if for some $i \leq n$,
$f(v_i)\not=p_1$,
then $p_2=f(v_{i_1})$ where $i_1$ is the least such  $i$,
and so on up to $p_k = f(v_n)$ (without loss
of generality, either $k=1$ or $p_k\not= p_1$).

A {\em reflex corner} is a triple
$p_{\ell-1},p_\ell,p_{\ell+1}$
of adjacent corners which are collinear, with
$p_{\ell-1}$ and $p_{\ell+1}$ on the same
side of $p_\ell$. (Interpret $p_{k+1}$ as $p_1$.)
\end{definition}

\begin{figure}
\centerline{\includegraphics[height=1.5in]{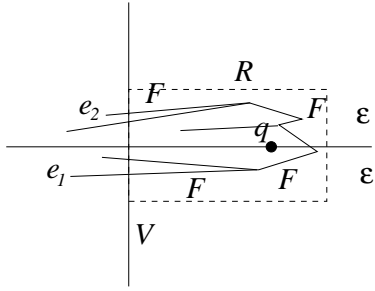}}
\caption{Why $s_1$ cannot be below $s_2$.}
\label{switchback.fig}
\end{figure}

\begin{figure}
\centerline{\includegraphics[height=1.5in]{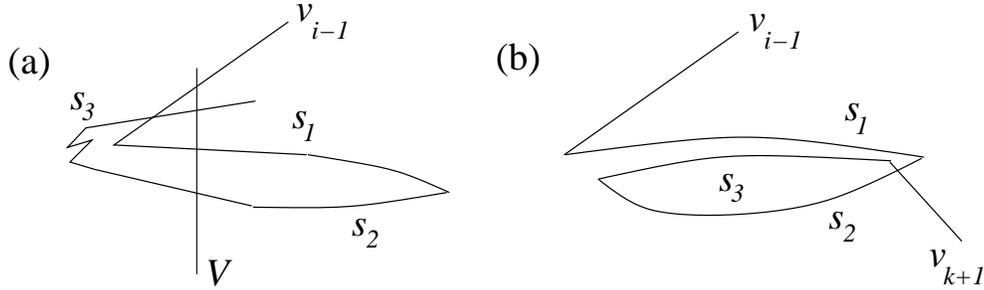}}
\caption{(a) $s_3$ is below $s_1$; (b) $v_k$ is between
$s_1$ and $s_2$.}
\label{s1s3.fig}
\end{figure}

Next we show that reflex corners do not exist.  Intuitively,
if $f(\partial F)$ made a $180^\circ$ turn
at $p_\ell$, then $f(F)$ would either be trapped in the line
$p_\ell p_{\ell+1}$ or it would surround it. The latter is
impossible since $p_\ell$ is a weighted average of neighbours
(Figure \ref{switchback.fig}).
This means that a sequence of monotone paths spirals inwards,
and the first edge to leave the line $p_\ell p_{\ell+1}$ crosses the spiral
(Figure \ref{s1s3.fig}).

\begin{lemma}
\label{lem: no reflex corners}
If $G$ is biconnected, $F$ a face,
and $f(\partial F)$ is not collinear, then
there are no reflex corners on $f(\partial F)$.
\end{lemma}

{\bf Proof.}
Suppose otherwise.
Let $S=v_i\ldots v_k$ be the longest subpath of $\partial F$
such that $p_{\ell-1},p_\ell,p_{\ell+1}$ is part of $f(S)$
and all of $f(S)$ is collinear.
Since $f(\partial F)$ is not contained in a line,
$S$ is a proper subpath of $\partial F$. Let
$I=f(S)$.  $I$ is a nondegenerate closed line-segment.
{\bf Claim} $I=f(v_i)f(v_k)$.

By definition of $S$, $f(v_{i-1})$ is not collinear with
$I$. If
$f(v_i)$ were interior to $I$ then either for some other
edge $e\in \partial F$
the edge $f(v_{i-1})f(v_i)$ would meet the relative interior of $f(e)$
transversally, which is impossible
(Lemma \ref{lem: vertex not interior edge}), or there
would be three or more edges $e$ in $\partial F$ such
that $f(e)$ was nondegenerate and met $f(v_i)$, not
all collinear, which is impossible
(Lemma \ref{lem: face 3 edges then collinear}).  The same arguments
apply to $v_k$. Thus $v_i$ and $v_k$ are
endpoints of $I$. Therefore either
$I=f(v_i)f(v_k)$,
or $f(v_i)=f(v_k)=p$, and for some other corner $q$,
$I=pq$.

The latter is impossible since both
edges
$f(v_{i-1})f(v_i)$ and $f(v_k)f(v_{k+1})$ would be
incident to $p$, and a third edge in $\partial F$,
mapped into $I$, would be incident to $p$, and
they would not be collinear, contradicting Lemma
\ref{lem: face 3 edges then collinear}.
Hence $I=f(v_i)f(v_k)$, as claimed.

We may assume that $I$ is contained in the $x$-axis with $f(v_i)$ left
of $f(v_k)$. Also, without loss of generality, we may
assume that for all sufficiently small $\delta$,
$f^\delta(F)$ is to the right of $f^\delta(v_j)f^\delta(v_{j+1})$
for $i-1 \leq j \leq k$
(Lemma \ref{lem: no flips}).
If it is not, rotate the
coordinate system through $180^\circ$.

Let  $s_1=v_i,\ldots, v_{i_1}$ be a maximal monotone path
(with respect to the $x$-axis), then let
$s_2 = v_{i_1},\ldots, v_{i_2}$ be a maximal
monotone path (in the other direction),
and continue until all of $v_i\ldots v_k$
has been subdivided into $m$ monotone paths.
Since
$p_{\ell-1} p_\ell p_{\ell+1}\subseteq f(S)$, $S$ is
not monotone, so $m\geq 2$.

{\bf Claim:} $s_1$ is above $s_2$.
Let $\{v_{r-1}, v_r\}$ be the last nondegenerate
edge in $s_1$ and $\{v_t,v_{t+1}\}$ the first in $s_2$:
$f(v_r) = f(v_t)$.
Let $q=f(v_r)=f(v_t)$.

Choose a vertex-avoiding vertical
line $V$ which intersects the interiors of $f(v_{r-1})q$ and
$qf(v_{t+1})$.

For every $\varepsilon>0$ there exists a $\delta > 0$
such that for all vertices $v$, $|f^\delta(v) - f(v)| < \varepsilon$,
and $V$ is $\delta$-vertex-avoiding. Let
$q_1$ and $q_2$ be the points where $V$ intersects the interiors of
$f^\delta(v_{r-1})f^\delta(v_r)$ and $f^\delta(v_t)f^\delta(v_{t+1})$.
We want to show that $q_1$ is above $q_2$.  Suppose otherwise,
so $q_1$ is below $q_2$.

Suppose $V=\{(a,y):~ y\in \IR\}$.

There is a topological sub-path $\pi$ of $f^\delta(s_1\cup s_2)$ joining
$q_1$ to $q_2$ and, since $f^\delta(s_1)$ and $f^\delta(s_2)$
cross $V$ from left to right and right to left respectively, and $V$ is
$\delta$-vertex-avoiding, $\pi$ is contained in the half-plane
$x \geq a$. By choice of $\varepsilon$, $\pi$ is
contained in the strip $-\varepsilon \leq y \leq \varepsilon$
and also in the open half-plane $ x < b+\varepsilon$,
where $q=(b,0)$.

Thus $\pi \subseteq R_{\varepsilon}$ where $R_{\varepsilon}$ is the rectangle
$$ x < b+\varepsilon, -\varepsilon < y < \varepsilon .$$  For
any edge $e$ incident to any vertex on this path,
$$ f^\delta (e) \cap \{(x,y):~ x\geq a\} \subseteq R_\varepsilon.$$  Allowing
$\varepsilon \to 0$,
we deduce that for every such edge $e$, $f(e)$
lies in the $x$-axis and its right-hand end is $q$. See
Figure \ref{switchback.fig}.
In particular, $v_r$ must be an internal vertex,
since every corner of $P$ has non-collinear incident edges.

Since $f(v_{r-1})$ is left of $q$, so is $f(v_r)$
(Lemma \ref{lem: one-sided}).
This is a contradiction: $s_2$ is below $s_1$, as claimed.
Similarly $s_3$ is above $s_2$, $s_4$ below $s_3$, and so on.

{\bf Claim:} for $3 \leq h \leq m$, $s_h$ is
below $s_1$ and above $s_2$.
To begin with, let $e_2$ and $e_3$ be the leftmost
nondegenerate edges occurring
in $s_2$ and $s_3$ (last and first, respectively). Since
$f(s_1)$ contains the leftmost point $f(v_i)$,
and the rightmost points in $f(s_1)$ and $f(s_2)$
are the same, $f(e_2)$ and $f(e_3)$ are contained
within $f(s_1)$. Also, $e_3$ is above $e_2$.
It follows that there exists a nondegenerate
edge $e_1$ in $s_1$ and a vertical vertex-avoiding
line $V$ which intersects $f(e_1),f(e_2),$ and $f(e_3)$.
Suppose that $s_3$ is above $s_1$.

Choose $\varepsilon > 0$ so that $V$ is
$\varepsilon$-vertex-avoiding.  Let $q_2$ be the
intersection of $f^\varepsilon(e_2)$ with $V$, and similarly $q_3$.
By hypothesis (and Lemma \ref{lem: asymmetric}),
$f(e_1)$ crosses $V$ between $q_2$ and $q_3$.
There is a topological path $\pi\subseteq f^\varepsilon(s_2\cup s_3)$
joining $q_1$ to $q_3$ which can be completed along $q_3q_1$ to
a Jordan curve $J$ which is crossed by $f^\varepsilon(e_1)$.
The left endpoint $p$ of $f^\varepsilon(s_1)$ is inside $J$.
$J,$ and $p$, can be made arbitrarily close to the $x$-axis,
and $f^\varepsilon(v_{i-1})f^\varepsilon(v_i)$ connects $p$ to a point bounded
away from the $x$-axis, so if $\varepsilon$ is small enough
then $f^\varepsilon(v_{i-1})f^\varepsilon(v_{i})$
crosses $\pi$, which is false.
Therefore $s_3$ is between $s_1$ and $s_2$.
See Figure \ref{s1s3.fig}.

If $s_4$ exists, then the right endpoints of
$f(s_3)$ and $f(s_4)$ coincide, and it follows
easily that $s_4$ is between $s_1$ and $s_2$. Generally speaking,
if $s_g$ exists, and
$g$ is odd (respectively, even), then
the argument concerning $s_3$ (respectively, $s_4$)
applies to show $s_g$ is between $s_1$ and $s_2$.
It follows that $f^\varepsilon (v_k)$ is
between $f^\varepsilon(s_1)$ and $f^\varepsilon(s_2)$ for
sufficiently small $\varepsilon$, and
$f^\varepsilon(v_k) f^\varepsilon (v_{k+1})$ crosses
$f^\varepsilon (s_1\cup s_2)$, which is impossible.
This contradiction shows that no reflex corner exists. {\bf Q.E.D.}\medskip

\begin{corollary}
\label{cor: convex polygon}
If $G$ is biconnected and $F$ is a face of $G$ then $f(\partial F)$ is
either a point, or a line-segment, or a convex polygon.
\end{corollary}

{\bf Proof.}
Let $S=f(\partial F)$ be described in the usual way as
a union of line-segments $p_i p_{i+1}$, $1 \leq i \leq k$
(interpret $p_{k+1}$ as $p_1$).  Suppose that not all points
$p_i$ are collinear.

Claim that $S$ is a simple polygon (though
adjacent line-segments $p_{i-1}p_i$ and $p_ip_{i+1}$ may
be collinear).  As usual, since $S$ is the limit
of simple polygons, it is connected, and edges do not
cross though they may overlap.

The interiors of no two
edge-images $p_i p_{i+1}$ and $p_j p_{j+1}$ can overlap.
Otherwise one can extend them to two maximal collinear chains
of edges which overlap.  These chains contain no reflex corners
(Lemma \ref{lem: no reflex corners}).
Let $I$ be their intersection.
$I$ is bounded by points $p$ incident to the images of
three or more edges, not all collinear, which is impossible
(Lemma \ref{lem: face 3 edges then collinear}): this
proves that edge images do not overlap.

Again, if a point $p$ is incident to the images of more than
two edges, then all these edge-images are collinear
(Lemma \ref{lem: face 3 edges then collinear}), and edge
images would overlap, which is false. Therefore $S$ is a simple polygon
(though successive edges could be collinear).

It remains to show that $S$ is a convex polygon.
Otherwise it has a concave corner
$p_{\ell-1}p_\ell p_{\ell+1}$ in the sense that
the interior of $S$ is on the concave side
of this broken line. In particular, $p_\ell$
is interior to the convex hull of $S$ so $p_\ell$
is not a corner of the bounding polygon $P$.

By the argument
showing that reflex corners do not exist, as illustrated in
Figure \ref{switchback.fig}, there would exist a
vertex $v$ such that $f(v) = p_\ell$, for all
neighbours $u$ of $v$, $f(u)$ is in the convex
wedge containing $p_{\ell-1},p_\ell,$ and $p_{\ell+1}$,
and for some neighbour $u$ of $v$, $f(u)\not= f(v)$.
Also, $v$ is an internal vertex.
This contradicts Lemma \ref{lem: one-sided}.
{\bf Q.E.D.}\medskip

\begin{lemma}
\label{faces dont collapse} 
If $G$ is convex embeddable then
the map $f$ does not collapse faces onto nondegenerate line-segments.
\end{lemma}

{\bf Proof.}
For $f$ to collapse a face $F$ into a nondegenerate line-segment
means that $f(\partial F)$ is not a point and is contained
in a line $L$.
Suppose this is the case.
$F$ must be bounded.  Let $I$ be the maximal connected union of
nondegenerate line-segments, including
$f(\partial F)$, which are collinear and are
the images of face-boundaries.

Let $V$ be a vertex-avoiding directed line orthogonal to $L$
which intersects the relative interior of $f(\partial F)$,
and is directed into $\hull(P)$ (this only matters
if $I\subseteq P$). Therefore $V$ intersects
at least one edge-image above $L$.
Let $e_1'$ be the highest edge (with respect to
the relation `$e_1$ is above $e_2$ on $V$'
(\ref{def: monotone path}))
such that $f(e_1')\subseteq L$ and $V\cap f(e_1')\not=\emptyset$.
Let $e_1$ be the lowest edge above $e_1'$ along $V$.

Choose $\varepsilon>0$ so that $V$ is $\varepsilon$-vertex-avoiding. For
all $\delta$ with $0<\delta\leq \varepsilon$, $V\cap f^\delta(F)$
is nonempty.

Also, $V\cap f^\varepsilon(e_1)$
and $V\cap f^\varepsilon(e_1')$ are joined along
$V$ by a line-segment which meets no other edge-image.
Therefore they are in the
same face of $f^\varepsilon(G)$ and hence
there exists a (bounded) face $F_1$ of $G$ containing
both $e_1$ and $e_1'$. Since $f(F_1)$ intersects
$f(e_1)$ and $f(e_1')$, $f(\partial F_1)$
is a convex polygon $S_1$ joining points $p_i$, some of which
may be collinear, but which are in cyclically
monotone order around $S_1$
(Corollary \ref{cor: convex polygon}).
$S_1\cap L$ is a line-segment $I_1$.  Let
$P_1 = u_1,\ldots,v_1$ be the maximal path such that
$f(P_1)=I_1$.
$P_1$ contains $e_1'$ and
its complementary path $Q_1\subseteq \partial F_1$, joining $u_1$ to $v_1$,
contains $e_1$.

There are two cases: (i) $I$
intersects the interior of $\hull(P)$
and (ii) $I$ is contained in a side of $P$.

In case (i), if we reverse the direction of
$V$, we get corresponding data
$e_2', e_2, F_2, S_2, P_2, u_2, v_2, $ and $Q_2$. We shall
see that $\partial F_1 \cap \partial F_2 $ is disconnected,
so $G$ is not convex embeddable.

Without loss of generality, $L$ is the $x$-axis,
$f(u_1)$ is left of $f(v_1)$, and $f(u_2)$ is left of $f(v_2)$.

First, for all sufficiently small $\delta$,
$V\cap f^\delta(e_1')\not= V\cap f^\delta(e_2')$.
This is because $f(e_1)\cap V$ and $f(e_2)\cap V$ are on
opposite sides of $L$, so for all sufficiently
small $\delta$, $V\cap f^\delta(e_1)$ and $V\cap f^\delta(e_2)$
are on opposite sides of $L$ and their distance from
$L$ is bounded below, whereas $f^\delta(F)$ can
be made arbitrarily close to $L$.  Therefore
$V$ intersects $f^\delta(F)$ between
$f^\delta(e_1)$ and $f^\delta(e_2)$.  By choice
of $e_1'$ and $e_2'$, $V\cap f^\delta(F)$ separates
$V\cap f^\delta(e_1')$ from $V\cap f^\delta(e_2')$.
Hence the intersection-points differ.

Since $f(\partial F_1)$ is a convex polygon (Corollary
\ref{cor: convex polygon}) and $V$ is $\varepsilon$-vertex-avoiding and
intersects $f(e_1)$ and $f(e_1')$,
$V$ intersects $f^\varepsilon(\partial F_1)$ in these edges
alone. Hence $V\cap f^\varepsilon(P_1)=V\cap f^\varepsilon(e_1')$.
Also $V\cap f^\varepsilon(P_2)=V\cap f^\varepsilon(e_2')$.
Therefore $P_1\not= P_2$.

\begin{figure}
\centerline{\includegraphics[height=1.5in]{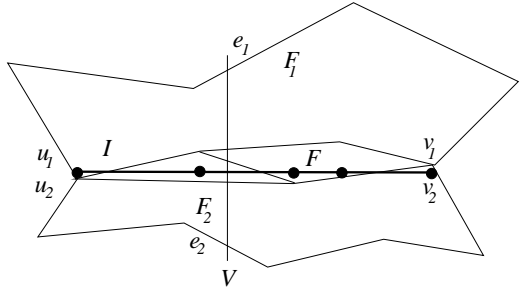}}
\caption{Illustrating $F_1$ and $F_2$ under $f^\delta$
(Lemma \protect\ref{faces dont collapse}). Note:
$u_1=u_2$ and $v_1 = v_2$.}
\label{collaps1.fig}
\end{figure}

Next, $f(u_1) = f(u_2)$. Otherwise, without
loss of generality, $f(u_1)$ is
in the relative interior of $f(u_2)f(v_2)$. Thus
$f(u_1)$ is inside $P$ and $u_1$ is an
internal vertex. Since
$u_1$ has a neighbour $w$ in $\partial F_1$ where
$f(w_1)\notin L$, and $f$ is a convex combination map,
$u_1$ has a neighbour $y_1$ such that $f(y_1)$ and $f(w_1)$
are on opposite sides of $L$. Then the line-segment
$f(u_1)f(y_1)$ intersects the interior of $S_2$.
Therefore, for sufficiently small $\delta>0$,
$f^\delta(u_1)f^\delta(y_1)$ intersects the interior
of the face $f^\delta(F_2)$, which is impossible.
From this contradiction, $f(u_1) = f(u_2)$, and
also $u_1$ has neighbours $w_1$ and $y_1$
such that $f(w_1)$ and $f(y_1)$ are on opposite
sides of $L$; similarly, $u_2$ has neighbours
$w_2$ and $y_2$ with $f(w_2)$ and $f(y_2)$
on opposite sides of $L$.
See Figure \ref{collaps1.fig}.

Next, $u_1 = u_2$.
If $u_1\not= u_2$ then there are two distinct paths
$s_1=w_1 u_1 y_1$ and $s_2=w_2 u_2 y_2$  such
that $f(s_1)$ crosses $f(s_2)$. For sufficiently
small $\delta>0$, $f^\delta(s_1)$ would
cross $f^\delta(s_2)$, which is impossible.
Hence $u_1=u_2$.  Similarly, $v_1=v_2$.

Thus $\partial F_1 \cap \partial F_2$ contains
$u_1$ and $v_1$. If $\partial F_1\cap \partial F_2$
is connected then it contains a path $Q$
joining $u_1$ to $v_1$ in both $\partial F_1$
and $\partial F_2$, $Q=P_1$ or $Q=Q_1$,
and $Q=P_2$ or $Q=Q_2$. But $Q_1$ contains
$e_1\notin \partial F_2$, so $Q\not= Q_1$;
also, $Q\not= Q_2$.  Therefore $P_1=P_2$
which has already been shown to be false, so
$\partial F_1\cap \partial F_2$ is disconnected. This
concludes Case (i).

Case (ii): $I$ is contained in a side of $P$.
Let
$H$ be the closed half-plane containing $P$ and
bounded by $L$.
We have the data
$V,e_1', e_1, F_1, S_1, P_1, u_1, v_1, $ and $Q_1$.
First, $f(u_1)$ is a corner of $P$.  Otherwise
$u_1$ is an internal vertex,
and since all vertices are
mapped into $H$, and $f(v)\notin L$ where $v$ is the neighbour of
$u_1$ in $Q_1$, this contradicts Lemma
\ref{lem: one-sided}.
Since $f(u_1)$ is a corner, there is only
one vertex mapped to $f(u_1)$
(Lemma \ref{injective on external vertices}),
so $u_1,$ and similarly $v_1$, is an external vertex.
Let $e_2' = \{u_1,v_1\}$, so $f(e_2')=I$.
$V\cap f(e_1)$ is bounded away from $L$ and
and $f(\partial F)\subseteq L$, so for all sufficiently
small $\delta$, $V\cap f^\delta(F)$
is between $V\cap f^\delta(e_1')$
and $V\cap f^\delta(e_2')$. Therefore $e_2'$ is not
incident to $\partial F_1$,
whereas $u_1, v_1 \in \partial F_1$, and $G$ has
an inverted subgraph, which is false.
{\bf Q.E.D.}\medskip

\begin{corollary}
\label{cor: edges dont overlap}
If $G$ is convex embeddable and $e\not=e'$ are edges then
$f(e)$ and $f(e')$ don't overlap.
\end{corollary}

{\bf Proof.}
Otherwise take a directed vertex-avoiding line $V$ intersecting
$f(e)\cap f(e')$ orthogonally.  Without loss
of generality, $e$ is above $e'$ along $V$.
Let $F$ be the face incident to $e$ such that $f^\delta(F)$ 
is below $f^\delta(e)$ for all sufficiently small $\delta$.
$f^\delta(F) \cap V$ is between $f^\delta(e)$ and
$f^\delta(e')$, so in the limit
$f(\partial F)$ is not a point nor a simple polygon, so it is
a nontrivial line-segment (Corollary \ref{cor: convex polygon}),
which is impossible. {\bf Q.E.D.}\medskip

\begin{lemma}
\label{edges dont collapse} 
If $G$ is convex embeddable, then $f$ does
not collapse edges to points.
\end{lemma}

{\bf Proof.} (This is similar to Lemma \ref{lem: one-sided}.)
Otherwise let $H$ be a maximal connected
subgraph of $G$ such that $f(H)$ is a single point,  $p$, say.
For each $u\in H$, let $N_u$ be the set of neighbours $v$
of $u$ such that $f(v)\not=p$.
There must be more than one
vertex $u$ such that $N_u\not= \emptyset$, since
otherwise $G$ or some $G\backslash u$ would be disconnected.

Given $u_1\not= u_2\in H$, $v_i,w_i\in N_{u_i}$, $i=1,2$,
the paths $f(v_1)f(u_1)f(w_1)$ and $f(v_2)f(u_2)f(w_2)$
cannot cross, since otherwise, for some $\delta>0$,
$f^\delta(v_1)f^\delta(u_1)f^\delta(w_1)$
and $f^\delta(v_2)f^\delta(u_2)f^\delta(w_2)$ would cross.

By Lemma \ref{injective on external vertices}, all vertices
in $H$ are internal.
Let $D$ be a closed disc centred at $p$ such that for every
vertex $v$, if $f(v)\not=p$, then $f(v)\notin D$.  We can partition
$\partial D$ into minimal arcs $A_u$, one for each $u$ in $H$
such that $N_u\not= \emptyset$, where
$$ A_u\supseteq \partial D \cap \{pf(v):~ v\in N_u\}.$$ By
Lemma \ref{lem: one-sided}, there are exactly two such
arcs $A_{u_1}$ and $A_{u_2}$, disjoint except perhaps at their endpoints,
and for all $v\in A_{u_1}\cup A_{u_2}$, $pf(v)$ are collinear,
and also $u_1$ has neighbours $v_1$ and $w_1$ in $N_{u_1}$
such that $pf(v_1)$ and $pf(v_2)$ do not overlap.  The
same goes for $u_2$.  It follows that there must be overlapping
edges $p f(v_1)$ and $pf(v_2)$, say, contradicting Corollary
\ref{cor: edges dont overlap}.
{\bf Q.E.D.}\medskip

We have established that if $G$ is convex embeddable then
$f$ maps face boundaries
injectively to convex polygons.
This is enough to prove that $f$ is an embedding,
by Tutte's arguments \cite{tutte}, which are as follows.

Provisionally, let
us define $f(F)$ as $f(\partial F) \cup \int (f (\partial F))$
for every bounded face $F$.

\begin{figure}
\centerline{\includegraphics[height=1in]{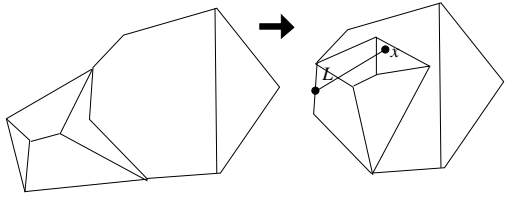}}
\caption{loosely illustrating $G$ and a many-to-one map
which is one-to-one on individual faces.}
\label{covnum.fig}
\end{figure}

For every point $x$ inside the
bounding (convex) polygon $P$, its {\em covering number} is the number
of faces $F$ such that $x \in f(F)$. See Figure \ref{covnum.fig}.

This number is 1 on the bounding polygon, and if we take
a vertex-avoiding line $L$ from the boundary to $x$, the number
can only change where an edge is crossed.  However,
to every internal edge $e$ there are exactly
two incident faces $F_1$ and $F_2$, and $f(F_1)$
and $f(F_2)$ are incident to $f(e)$ from opposite sides.
Otherwise $f^\delta(F_1)$ and $f^\delta(F_2)$ would
overlap for sufficiently small $\delta$.
It follows
that the covering number does not change 
where $L$ crosses edges, so it is $1$ for all $x$,
and $f$ is injective.  This completes the proof of
our main theorem.

\begin{theorem}
\label{embedding theorem} 
If $G$ is convex embeddable then $f$ is an embedding.
Therefore $G$ admits
a convex embedding if and only if every
convex combination map is an embedding.
\qed
\end{theorem}

\section{Ambient isotopy}
\label{ambient isotopy section} 
In \cite{stein}, Stein considered plane embedded graphs in which
every face boundary is a simple cycle and no two {\em bounded} faces have
disconnected intersection
(see also \cite{tutte60}).
By our earlier results, all nodally
3-connected plane embedded graphs have this property.
Stein showed that all such graphs admit convex embeddings, where
the bounded faces map to convex polygons, so long as edges can
be embedded piecewise linear rather than straight.
Equivalently,
one can allow new vertices (of degree 2) to be introduced.
The existence of inverted subgraphs becomes irrelevant.
Let us call such graphs {\em general convex embeddable},
or GCE for short.  Stein also allowed them to have multiple edges.

Stein remarked in \cite{stein} that any two (convex) embeddings, with the same
orientation, of a GCE graph are ambient isotopic,
but does not include a proof.

\begin{definition}
\label{isotopy} 
Given topological spaces $X$ and $Y$, an {\em isotopy} is a continuous map
$h: [0,1]\times X \to Y$ such that for each $t,$ $0\leq t \leq 1,$
the map $h_t: X\to Y;~~ x\mapsto h(t,x)$ is a homeomorphism.
\end{definition}

This section gives an outline proof of
the following isotopy theorem
(Corollary \ref{isotopy theorem}).
Let $G^1$ and $G^2$
be two plane embeddings
of the same GCE graph $G$, such that
their external boundaries are images of the same cycle
$C$ of $G,$ with the same orientation.
Then there exists an isotopy: $\IR^2\to \IR^2$
taking the vertices, edges, and faces of $G^1$ to those of $G^2.$

\begin{proposition}
\label{can merge countries} 
Suppose $G$ is a GCE plane embedded graph
Then
either $G$ has just one bounded face or there exist two bounded
faces $F'$ and $F''$ such that $\partial F' \cap \partial F'' = Q$
is nonempty (and connected), and if
$F = F' \cup \int(Q)\cup F'',$ then
for every other face $A$ of $G$, $\partial A \cap\partial F$ is
connected.

Furthermore, if $G'$ is the embedded graph obtained by
removing the edges and inner vertices on $Q,$ hence
merging $F'$ and $F''$ into a single face $F,$ then $G'$ is also GCE,
with the same external boundary as $G.$
(The first part was proved in {\rm \cite{stein}}, and the rest
follows immediately.)\qed
\end{proposition}

\begin{definition}
\label{ambient isotopic defined} 
Let $G^1$ and $G^2$ be two plane embedded graphs.  The
embeddings are {\em ambient homeomorphic}
(respectively, {\em ambient isotopic}) if there is
a homeomorphism (respectively,
an isotopy) from $\IR^2$ to itself taking the
vertices, edges, and faces of $G^1$ bijectively
onto those of $G^2.$
\end{definition}

\begin{definition}
\label{theta graph} 
A {\em $\theta$-graph} is a plane embedded graph consisting
of two nodes connected by three disjoint paths.
It resembles the Greek letter $\theta$.
\end{definition}

\begin{lemma}
\label{theta graph homeomorphism} 
If $G^1$ and $G^2$ are plane embeddings of a
$\theta$-graph $G$, with the same orientation,
then they are ambient homeomorphic.
(Follows from the Sch\"onflies theorem
\ref{schoenflies theorem}: proof omitted.)\qed
\end{lemma}

\begin{corollary}
\label{ambient homeomorphic} 
If $G^1$ and $G^2$ be two GCE embeddings
of the same graph $G$ with the same orientation
and the same boundary cycle,
then they are ambient homeomorphic.
\end{corollary}

{\bf Proof.}  This is a simple application of
Stein's result (Lemma \ref{can merge countries}), and
is by induction on the number of bounded faces. If
$G$ is a simple  cycle then this is just
the Sch\"onflies Theorem (Proposition \ref{schoenflies theorem}).

For the inductive step, choose faces $F'$ and $F''$ of
$G^1$ separated by a path $Q$ such that $F=F'\cup \int(Q)\cup F''$
has the properties stated in Lemma \ref{can merge countries}.
Let $H$ be the subgraph of $G$ obtained by removing the
edges and inner vertices of $Q,$ and let $H^1$ be
the modified embedding where $F'$ and $F''$ are merged into $F.$
Then $H^1$ is a GCE embedding of $H.$
Similarly a modified embedding $H^2$ is obtained from $G^2.$
By induction, $H^1$ and $H^2$ are ambient homeomorphic through
a homeomorphism $h'$.  Let
$D^1$ and $D^2$ be the images of $\overline{F}$ under
the respective embeddings.  $D^2 = h'(D^1)$.
They contain images
$Q^1$ and $Q^2$ of the path $Q.$

By Lemma \ref{theta graph homeomorphism},
there exists a homeomorphism $h:D^1\to D^2$
which agrees with $h'$ on $\partial D^1$ and takes
$(F')^1$ to $(F')^2$,
$(F'')^1$ to $(F'')^2$, and $Q^1$ to $Q^2$,
and also takes the vertices and edges in $Q^1$ to
those in $Q^2$.  Extend $h$ to $\IR^2$ by making
it coincide with $h'$ outside $(\partial F)^1.$
Then $h$  is an ambient homeomorphism between
$G^1$ and $G^2.$ {\bf Q.E.D.}\medskip

\begin{corollary}
\label{isotopy theorem} 
If $G^1$ and $G^2$ are GCE embeddings of the
same graph with the same external boundary in the
same anticlockwise order $C^1$ and $C^2$, then the embeddings are connected
by an isotopy.
\end{corollary}

{\bf Sketch proof.} There is an ambient homeomorphism $h$ connecting them
(Corollary \ref{ambient homeomorphic}). According to
\cite{stillwell}, $h$ is isotopic to the identity or to
reflection in the $x$-axis.  Furthermore,
if $h$ preserves the orientation of any Jordan
curve, as it does in this case, it is isotopic to the
identity.  This yields an isotopy carrying
$G^2$ to $G^1.$\qed\medskip

Let $G$ be a convex embeddable plane-embedded graph.
We can let $G^1$ correspond to the identity map on $\IR^2$,
and $G^2$ correspond to an orientation-preserving convex combination
map $f$.  Then

\begin{corollary}
\label{convex embeddable graph isotopic} 
Let $G$ be an convex embeddable graph
and suppose $f$ is an orientation-preserving convex-combination map.
Then there is an istotopy of $\IR^2$ taking
each vertex $v$, edge $e$, and face $F$ of $G$ to
$f(v)$, $f(e)$, and $f(F)$, respectively.\qed
\end{corollary}

\section*{Acknowledgement}

The author is grateful to
Colum Watt for useful information about isotopies.

\end{document}